\documentclass[trackchanges,twocolumn]{aastex701}

\usepackage{amsmath}
\usepackage{subfigure}
\usepackage{booktabs}
\usepackage[hang,flushmargin]{footmisc}
\usepackage{multirow}
\usepackage{threeparttable}
\usepackage{longtable}
\usepackage{graphicx}%
\usepackage{amssymb,amsfonts}%
\usepackage{mathrsfs}%
\usepackage{xcolor}%
\usepackage{textcomp}%
\usepackage{algorithm}%
\usepackage{algorithmicx}%
\usepackage{algpseudocode}%
\usepackage{listings}%
\usepackage{overpic}
\usepackage{makecell}
\usepackage{float}
\usepackage{hyperref}
\usepackage{placeins}
\usepackage{lineno}
\usepackage{natbib}
\usepackage{booktabs}

\begin{document}

%\usepackage{natbib}

%75 characters including spaces.
\title{EP251129a: A Fast X-ray Transient with Two X-Ray Plateaus and Its Off-axis Afterglow}% EP251129a: An Ultra-long Gamma-Ray Burst with Two X-Ray Plateaus and an unusual Radio Afterglow

% \correspondingauthor{Zi-Gao Dai (daizg@ustc.edu.cn); Tao An (taoan3068@gmail.com)}

\correspondingauthor{Zi-Gao Dai, Tao An}
\email{daizg@ustc.edu.cn, antao2008@ustc.edu.cn}

\author[orcid=0000-0002-5225-637X]{Yue Wu}
\affiliation{School of Astronomy and Space Science, Nanjing University, Nanjing 210023, China}
\affiliation{Key Laboratory of Modern Astronomy and Astrophysics (Nanjing University), Ministry of Education, Nanjing 210023, China}
\email{yue_wu@smail.nju.edu.cn}

\author[orcid=0000-0002-7351-5801]{Ailing Wang}
\affiliation{Yunnan Observatories, Chinese Academy of Sciences, 396 Yangfangwang, Guandu District, Kunming 650216, China}
\affiliation{Key Laboratory of Particle Astrophysics, Institute of High Energy Physics, Chinese Academy of Sciences, Beijing 100049, China}
\affiliation{State Key Laboratory of Radio Astronomy and Technology, National Astronomical Observatories, Chinese Academy of Sciences, Beijing 100101, China}
\email{ailing.wang.wal@gmail.com}

\author[orcid=0009-0002-0137-2650]{Senlin Pang}
\affiliation{Department of Astronomy, University of Science and Technology of China, Hefei 230026, China}
\email{forestpang@mail.ustc.edu.cn}

\author[0000-0003-0691-6688]{Yu-Han Yang}
\affiliation{Department of Physics, University of Rome ``Tor Vergata'', via della Ricerca Scientifica 1, I-00133 Rome, Italy}
\email{yuhan.yang@roma2.infn.it}

\author[orcid=0009-0002-0930-1830]{Biao Zhang}
\affiliation{Department of Astronomy, University of Science and Technology of China, Hefei 230026, China}
\email{biaozhang@mail.ustc.edu.cn}

\author[orcid=0000-0002-7835-8585]{Zi-Gao Dai}
\affiliation{Department of Astronomy, University of Science and Technology of China, Hefei 230026, China}
\email{daizg@ustc.edu.cn}

\author[0000-0003-4341-0029]{Tao An}
\affiliation{Department of Astronomy, University of Science and Technology of China, Hefei 230026, China}
\email{antao2008@ustc.edu.cn}

\author[orcid=0000-0001-9648-7295]{Jin-Jun Geng}
\affiliation{Purple Mountain Observatory, Chinese Academy of Sciences, Nanjing 210023, China}
\email{jjgeng@pmo.ac.cn}

\author[orcid=0000-0002-6299-1263]{Xue-Feng Wu}
\affiliation{Purple Mountain Observatory, Chinese Academy of Sciences, Nanjing 210023, China}
\email{xfwu@pmo.ac.cn}

\author[orcid=0000-0002-9615-1481]{Hui Sun}
\affiliation{National Astronomical Observatories, Chinese Academy of Sciences, Beijing 100101, China}
\email{hsun@nao.cas.cn}

\author[orcid=0000-0002-4304-2759]{Di Xiao}
\affiliation{Purple Mountain Observatory, Chinese Academy of Sciences, Nanjing 210023, China}
\email{dxiao@pmo.ac.cn}

\author[orcid=0000-0002-9037-8642]{Jia Ren}
\affiliation{Purple Mountain Observatory, Chinese Academy of Sciences, Nanjing 210023, China}
\email{renjia@pmo.ac.cn}

\author[orcid=0000-0002-2191-7286]{Chen Deng}
\affiliation{School of Astronomy and Space Science, Nanjing University, Nanjing 210023, China}
\affiliation{Key Laboratory of Modern Astronomy and Astrophysics (Nanjing University), Ministry of Education, Nanjing 210023, China}
\email{dengchen@nju.edu.cn}

\author[orcid=0000-0001-8278-2955]{Yehao Cheng}
\affiliation{South-Western Institute for Astronomy Research, Yunnan Key Laboratory of Survey Science, Yunnan University, Kunming, Yunnan 650504, People's Republic of China}
\email{yhcheng@mail.ynu.edu.cn}

\author[0000-0003-4157-7714]{Fa-Yin Wang}
\affiliation{School of Astronomy and Space Science, Nanjing University, Nanjing 210023, China}
\affiliation{Key Laboratory of Modern Astronomy and Astrophysics (Nanjing University), Ministry of Education, Nanjing 210023, China}
\email{fayinwang@nju.edu.cn}

\author[orcid=0000-0001-6374-8313]{Yuan-Pei Yang}
\affiliation{South-Western Institute for Astronomy Research, Yunnan Key Laboratory of Survey Science, Yunnan University, Kunming, Yunnan 650504, People's Republic of China}
\email{ypyang@ynu.edu.cn}

\author[0000-0002-1869-7817]{Eleonora Troja}
\affiliation{Department of Physics, University of Rome ``Tor Vergata'', via della Ricerca Scientifica 1, I-00133 Rome, Italy}
\email{eleonora.troja@uniroma2.it}

\author[0000-0002-9700-0036]{Brendan O'Connor}
\affiliation{McWilliams Center for Cosmology and Astrophysics, Department of Physics, Carnegie Mellon University, Pittsburgh, PA 15213, USA}
\email{boconno2@andrew.cmu.edu}

%%%%%%%%%%%%%%%

\author[orcid=0009-0003-9214-7316]{Wen-Jie Zhang}
\affiliation{National Astronomical Observatories, Chinese Academy of Sciences, Beijing 100101, China}
\affiliation{School of Astronomy and Space Science, University of Chinese Academy of Sciences, Chinese Academy of Sciences, Beijing 100049, People's Republic of China}
\email{zhangwj@bao.ac.cn}

\author[orcid=0009-0007-6104-337X]{Yuan Liu}
\affiliation{National Astronomical Observatories, Chinese Academy of Sciences, Beijing 100101, China}
\email{liuyuan@bao.ac.cn}

\author[orcid=0000-0001-8266-3024]{Weimin Yuan}
\affiliation{National Astronomical Observatories, Chinese Academy of Sciences, Beijing 100101, China}
\affiliation{School of Astronomy and Space Science, University of Chinese Academy of Sciences, Chinese Academy of Sciences, Beijing 100049, People's Republic of China}
\email{wmy@nao.cas.cn}

\author[orcid=0009-0004-7110-3450]{Yahui Jiang}
\affiliation{School of Astronomy and Space Science, Nanjing University, Nanjing 210023, China}
\affiliation{Key Laboratory of Modern Astronomy and Astrophysics (Nanjing University), Ministry of Education, Nanjing 210023, China}
\email{yahjiang@smail.nju.edu.cn}

\author[orcid=0000-0002-8385-7848]{Yun Wang}
\affiliation{Purple Mountain Observatory, Chinese Academy of Sciences, Nanjing 210023, China}
\affiliation{Department of Astronomy, University of California, Berkeley, CA 94720-3411, USA}
\email{wangyun@pmo.ac.cn}

\author[orcid=0000-0002-0170-0741]{Cuiyuan Dai}
\affiliation{School of Astronomy and Space Science, Nanjing University, Nanjing 210023, China}
\affiliation{Key Laboratory of Modern Astronomy and Astrophysics (Nanjing University), Ministry of Education, Nanjing 210023, China}
\email{cydai@smail.nju.edu.cn}

\author[orcid=0009-0001-8733-2088]{Hucheng Ding}
\affiliation{Department of Physics, Anhui Normal University, Wuhu, Anhui 241002, China}
\email{dinghc@ahnu.edu.cn}

\author[orcid=0009-0005-8541-5209]{Tao Wu}
\affiliation{Department of Physics, Anhui Normal University, Wuhu, Anhui 241002, China}
\email{taowu@ahnu.edu.cn}

\author[orcid=0009-0000-0015-9124]{Kangrui Ni}
\affiliation{Institute of Astrophysics, Central China Normal University, Wuhan 430079, People's Republic of China}
\email{nikangrui@mails.ccnu.edu.cn}

\begin{abstract}
Extragalactic fast X-ray transients (EFXTs) detected by the \emph{Einstein Probe} are extending the parameter space of relativistic stellar explosions toward softer and longer-lived transients.
Here we report EP251129a/GRB 251129A at z=0.505, which exhibits two X-ray plateaus: an early plateau lasting $\sim 10^{4}\,\rm s$ followed by a rapid decay with spectral softening, and a second plateau extending to 60 days, placing it at the extreme long-duration end of Gamma Ray Burst (GRB) plateaus. 
The early optical emission evolves independently of the X-ray, indicating a different origin; whereas the unusual radio afterglow shows a rapid rise of $F_\nu\propto t^{1.9-2.5}$ and a delayed peak at $\sim36-40$ days, providing strong evidence for an off-axis geometry.
We find that the early X-ray behavior resembles those of ultra-long GRBs (ULGRBs) and can be naturally explained by the internal dissipation from a newborn millisecond magnetar undergoing gravitational-wave-dominated spin-down before collapsing into a black hole. The early optical, radio and late X-ray emission are well described by an off-axis two-component jet afterglow with additional energy injection, possibly powered by fallback accretion onto black hole.
EP251129a therefore connects the emerging EFXT population with ULGRB-like central-engine activity and off-axis afterglow, offering a rare opportunity to probe both the evolution of central engine and jet structure in a single event.
\end{abstract}

          % Tianguan Einstein Probe

\section{Introduction}
Extragalactic fast X-ray transients (EFXTs) are soft X-ray emissions lasting from seconds to hours at cosmological distances \citep[e.g.,][]{Jonker2013, Quirola2022, Quirola2023}. 
The \emph{Einstein Probe} (EP), with its wide-field sensitivity, has greatly expanded this population \citep{Yuan2022EP,Yuan2025}.
Some soft EP EFXTs exhibit X-ray activity that is substantially longer than their gamma-ray counterpart, while in others the gamma-ray emission may be weak or undetected \citep[e.g.,][]{Sun2025, LiuY2025, LiWX2025, Dai2026, LiAn2026}. 
With the growing evidence, population studies suggest that a substantial fraction of these events are closely related to long Gamma Ray Bursts (GRBs) and massive-star explosions, as indicated by their redshift distribution, energetics, supernova associations, and GRB-like afterglows \citep{OConnor2025ApJL, Chen2026}, although the origin is likely diverse, and several events may account for jetted TDEs or remain debated \citep{LiDY2026, OConnor2025_250702B, OConnor2025_240408a}.
EP therefore provides a complementary view of relativistic stellar explosions whose central-engine activity and jet emission may be incompletely sampled by conventional gamma-ray monitors.

Long-lived X-ray emission is particularly valuable for probing central-engine activities \citep{ZhangB2006, Troja2007, GaoHe2017}. An extreme manifestation is found in ULGRBs, a rare class with exceptionally high-energy activities persisting $\gtrsim10^3$--$10^4 \, \rm s$ or longer \citep[e.g.,][]{Gendre2013, Levan2014, Fu2024}. The origin remains debated, with proposed interpretations involving extended massive-star progenitors, black hole (BH) fallback accretion, and activities of rapid-rotating newborn magnetars \citep{Levan2014, Wu2013, Greiner2015, Gao2016, Zou2021, Fu2024}.
X-ray plateaus provide an additional diagnostic of such long-lived activity, which have been observed in many GRB X-ray lightcurves \citep[e.g.,][]{Willingale2007}. 
A subset followed by rapid decay has been recognized as internal X-ray plateaus attributed to internal emission from a long-lived central engine; in the newborn magnetar scenario, it can be powered by dissipation of the spin-down wind, with the abrupt decay marking the collapse into a BH \citep[e.g.,][]{Troja2007, Rowlinson2013, Lv2014, Lv2018}. At later times, shallow afterglow plateaus may instead arise from continued energy injection into the external shock or refreshed shocks driven by slower ejecta catching up \citep[e.g.,][]{Rees1998, Sari2000, ZhangB2006, Zhao2020, Deng2026}. 
On the other hand, the late multi-wavelength afterglows could also probe the jet geometry, for an off-axis jet, a delayed afterglow is expected as the relativistic beaming cone widens into the line of sight \citep{Lazzati2018, Gill2018, Pang2024a, Pang2024b, YangYH2026}. 

Here we report EP251129a/GRB 251129A at $z=0.505$ \citep{GCN43249}, an EFXT detected by EP and Space Variable Object Monitor (SVOM)/ECLAIRs independently \citep{GCN42877,GCN42884}. It exhibits a distinctive combination of multi-wavelength properties: an early internal X-ray plateau lasting $\sim10^4$ s followed by a rapid decay with spectral softening; an early optical component evolving independently of the X-rays; a second shallow X-ray plateau extending to $\sim67$ days; and a rapid-rising radio afterglow, $F_\nu\propto t^{1.9-2.5}$, with a delayed peak at $36-40$ days. 
We show that the early X-ray behaviors are naturally interpreted in a newborn magnetar collapse scenario, whereas the optical, radio, and late X-ray emission trace an off-axis two-component afterglow with continued energy injection from the BH fallback accretion. EP251129a therefore provides a rare case in which long-lived central-engine emission and the afterglow can be separated within a single event, linking the emerging EFXT population with ULGRB-like prolonged engine activity and off-axis afterglows.

\section{OBSERVATIONS AND DATA REDUCTION} \label{sec:observation}

\begin{figure*}
    % \vspace{0.1cm}
    \centering
    \begin{overpic}[width=0.355\textwidth]{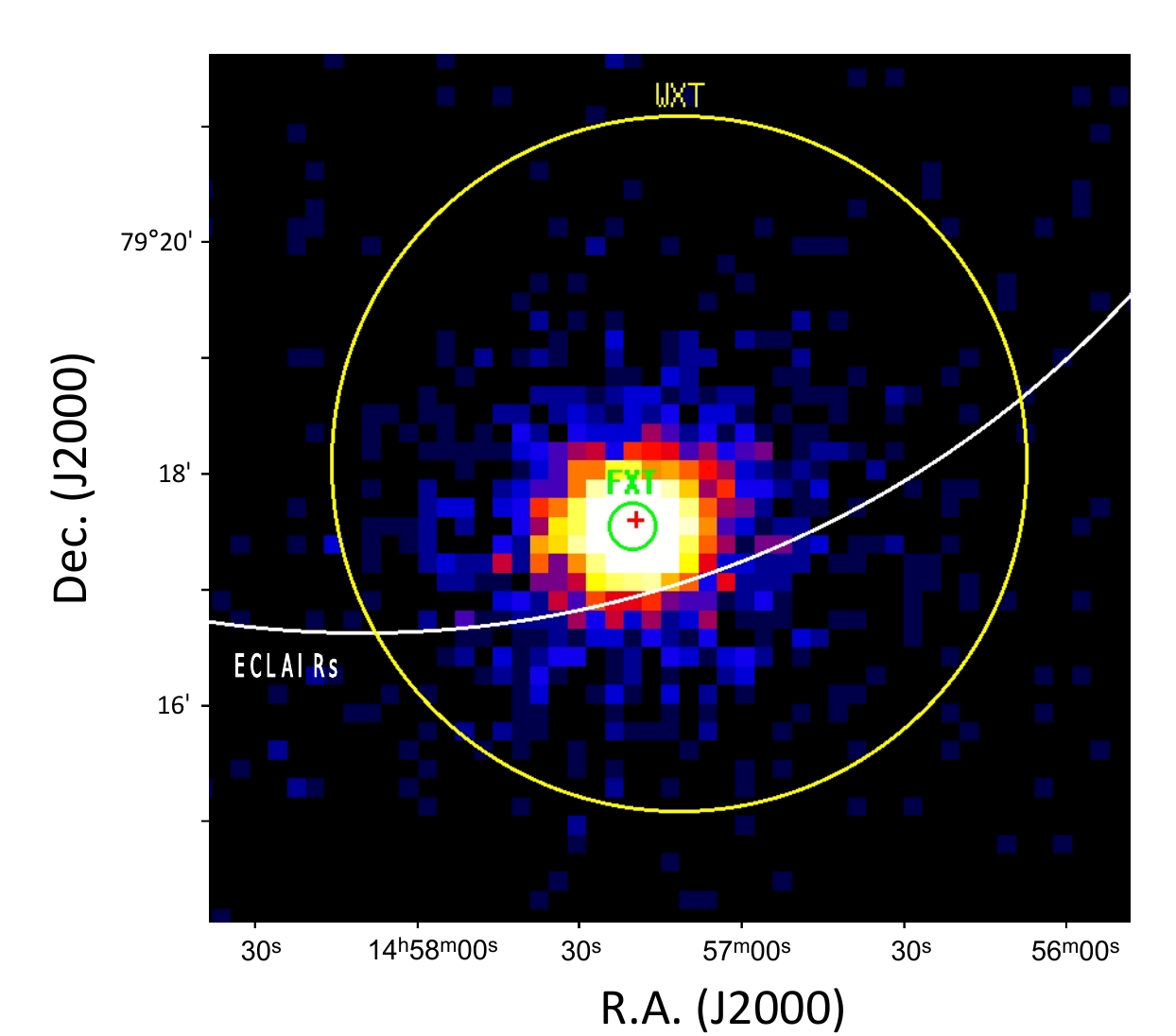}
        % \put(-1,103){\textbf{(a)}} % 标注
    \end{overpic}
    \hspace{0.01\textwidth} % 调整间距
    \begin{overpic}[width=0.379\textwidth]{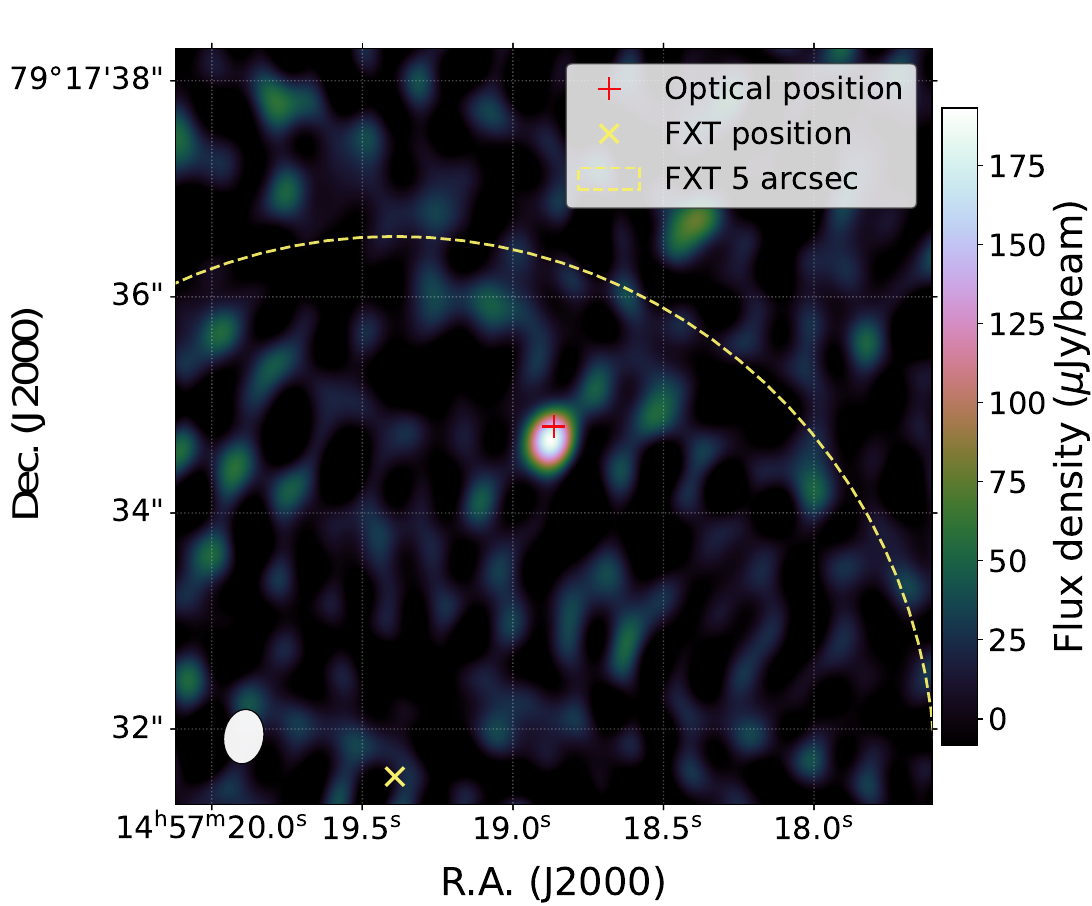}
      %   \put(-1,103){\textbf{(b)}} % 标注
    \end{overpic}
\caption{ %\textbf{X-ray and radio images of EP251129a.} 
\emph{Left}: EP-FXT image obtained during the first ToO observation. The part of large white circle represents the SVOM-ECLAIRs positional uncertainty with a radius of 8.96'. The yellow and green circles indicate the EP-WXT and EP-FXT positional uncertainty with a radius of 3.5' and 10''. The red cross denotes the optical position obtained from COLIBRÍ. 
\emph{Right}: VLA 11.5 GHz radio image of EP251129a observed on 2026 Feb 3. The optical and EP-FXT positions are marked by the red and yellow crosses. The yellow dashed circle denotes 5 arcsec radius from the FXT position. 
The synthesized beam is displayed as the white ellipse in the lower-left corner. The color bar gives the flux density in $\mu$Jy beam$^{-1}$.}
\label{fig:image}
\end{figure*}

\subsection{X-ray and high-energy observations}

The Wide-field X-ray Telescope (WXT) aboard \emph{Einstein Probe} (EP) detected EP251129a at 2025-11-29T03:14:23 UTC \citep{GCN42884}. The source is spatially and temporally coincident with GRB 251129A, triggered earlier by SVOM/ECLAIRs at $T_0=$2025-11-29T02:32:05 UTC \citep{GCN42877}. 
Subsequent ground refined analysis shows a detection from $T_0-41$ s to $T_0+3200$ s over 4--120 keV, with the time-averaged spectrum during the first interval ($T_0-41$ to $T_0+77$ s) well described by a power law of $\Gamma=1.7\pm0.1$ and $F_{4-120\,{\rm keV}}=(1.5\pm0.2)\times10^{-8} \,{\rm erg\,cm^{-2}\,s^{-1}}$  \citep{GCN42985}, and the later spectra become softer, $\Gamma\simeq2.0$--2.1, and connect smoothly to the long-lived X-ray emission subsequently observed by EP.

The Follow-up X-ray Telescope (FXT) aboard EP began autonomous follow-up approximately 100 s after the WXT trigger and localized the source at
R.A. = 224.3308 deg, Dec. = 79.2921 deg (J2000), with a 10 arcsec uncertainty (90\% confidence), consistent with the enhanced \emph{Swift} localization, the optical counterpart \citep{GCN42891,GCN42883}, and the VLA position as shown in Figure~\ref{fig:image}.
Subsequent EP-FXT ToO observations covered from $T_0+2.9$ hr to $T_0+107$ days, while the X-Ray Telescope (XRT; \citealp{Burrows2005}) onboard Neil Gehrels \emph{Swift} Observatory (\emph{Swift}) monitored from $T_0+4.5$ hr to $T_0+60$ days and detected it through $\sim33$ days. The observation log is given in Table~\ref{tab_FXT_XRT_obs}.

The EP-WXT data were processed over 0.5--4 keV band with the standard WXTDAS pipeline and the corresponding CALDB\citep{ChengHQ2025}. 
EP-FXT data were reduced with \texttt{fxtsoftware} v1.20 and the corresponding calibration files, following the standard reduction procedure \citep{ZhaoHS2025}.
And the \emph{Swift}-XRT data were processed with \texttt{HEASoft} v6.35.5 and the latest XRT CALDB.
The EP-FXT and \emph{Swift}-XRT spectra and light curves were analyzed over 0.5--10 keV.
Pile-up affects the first two EP-FXT observations and the first \emph{Swift}-XRT observation.
For the piled-up FXT-A and XRT data, central PSF was excluded and spectra were extracted from annular source regions.
FXT-B was severely more affected in full-frame mode and was therefore excluded from the spectral analysis of the first two EP-FXT observations.

The \emph{Fermi} Gamma-ray Burst Monitor (GBM; \citealp{Meegan2009}) had full spatial coverage around $T_0$, but neither an onboard trigger nor a consistent targeted-search signal was found between $T_0-50$ and $T_0+500$ s over timescales from 64 ms to 32.768 s.
Assuming the power-law spectrum measured during the first ECLAIRs interval ($\Gamma=1.7$), we derive a 3$\sigma$ 10--1000 keV flux upper limit of$2.1\times10^{-8}\ {\rm erg\,cm^{-2}\,s^{-1}}$for a 32.768 s interval.
This limit is moderately below the nominal extrapolation of the ECLAIRs spectrum, $3.3\times10^{-8}\ {\rm erg\,cm^{-2}\,s^{-1}}$, suggesting a possible spectral steepening or a high-energy break above the ECLAIRs band.
To test whether the late X-ray emission was intrinsic or from a nearby, unrelated source blended within the FXT/XRT PSF, and constrain the subsequent evolution, we obtained a \emph{Chandra} DDT observation (PI: Y.-H. Yang; ObsID 32430) on 2026 July 28, at $\sim242$ days after $T_0$, with a deflared exposure of 9.94 ks. No source was detected within a $1'$ search radius around the transient position.
Using a $1''$ source aperture centered on the VLA localization and a $15''$--$30''$ background annulus, we obtain a 3$\sigma$\,count-rate upper limit of $5.8\times10^{-4}\,{\rm counts\,s^{-1}}$ following the method of \citet{Kraft1991}. Adopting the spectral model inferred from the late-time FXT data, this corresponds to an unabsorbed 0.5--10 keV flux upper limit of $1.6\times10^{-14}\,{\rm erg\,cm^{-2}\,s^{-1}}$, strongly disfavoring a persistent unrelated contaminant in the late FXT/XRT detections.

\subsection{Optical observations}

Optical photometry was compiled from public GCN Circulars, including \emph{Swift}/UVOT observations with $u$ filter \citep{GCN42959}, SVOM/VT observations in VT\_B and VT\_R bands \citep{GCN42887,GCN42972}, and ground-based OSN, SAO RAS and Mondy observations in $R/R_c$-band \citep{GCN42897,GCN42899,GCN43062}.
Late-time multiband imaging with the Gran Telescopio Canarias (GTC) at 9.2 and 27.2 days revealed an optical rebrightening, consistent with the emergence of an associated supernova. Spectroscopy at 28.1 days detected emission lines from the underlying galaxy at a common redshift of $z=0.505$ \citep{GCN43249}. We adopt this redshift throughout this work. 
All optical measurements were corrected for Galactic extinction in the corresponding bands along the line of sight \citep{Schlafly2011}, while no correction for host-galaxy extinction was applied.

\subsection{Radio observations}

We conducted multi-epoch radio follow-up of EP251129a with the enhanced Multi-Element Radio Linked Interferometer Network (e-MERLIN) and the Karl G. Jansky Very Large Array (VLA)  from $T_0+4.6$ days to $T_0+150$ days (PI: T. An).
The e-MERLIN observation (project RR20009) was carried out in C band (4.82--5.33 GHz) from 2025 Dec. 3 to 5. 
The data were processed with the standard e-MERLIN pipeline followed by calibration and imaging in \texttt{CASA} \citep{CASA2022}. No significant emission was detected at the transient position, yielding a reported 3$\sigma$ upper limit of $F_\nu<66~\mu{\rm Jy\,beam^{-1}}$ \citep{GCN43040}.
The VLA observed at seven epochs between 2025 December 15 and 2026 April 28 (projects 25B-403 and 25B-410), with the first five epochs covering 4--8 GHz and the final two extending to 1.3--11.5 GHz. 
The data were calibrated and imaged in \texttt{CASA} using standard interferometric procedures, with 3C147 as the flux-density and bandpass calibrator and J1153+8058 as the phase calibrator.
The source was unresolved in all epochs, and we therefore adopted the image-plane peak flux densities, with $1\sigma$ uncertainties given by the local image rms measured in source-free regions around the target. The resulting measurements are listed in Table~\ref{tab:vla1}.

\begin{figure}[t] 
    \centering
    \begin{overpic}[width=0.478\textwidth]{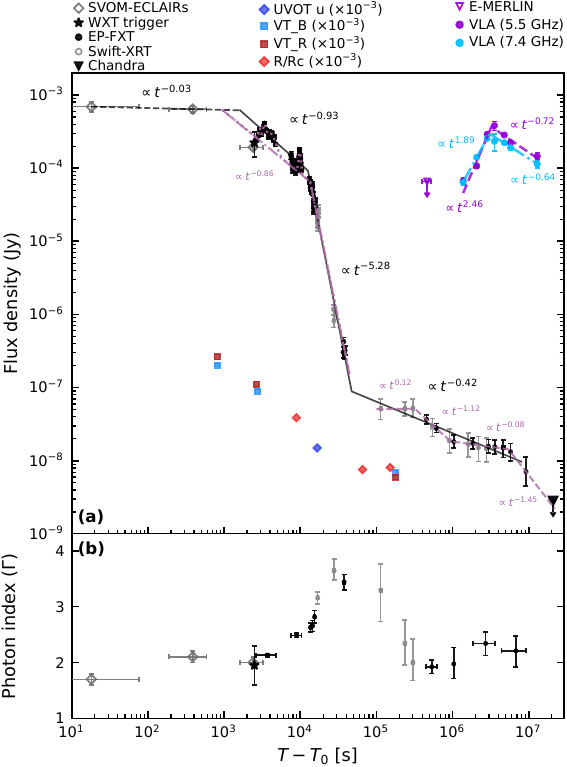}
    \end{overpic}
    % 总体标题
    \caption{
    \textbf{Multi-wavelength evolution of EP251129a.} \emph{Top}: X-ray flux density at 1 keV from SVOM/ECLAIRs, EP-WXT, EP-FXT, \emph{Swift}-XRT, and the \emph{Chandra} upper limit, together with optical (scaled for display) and radio measurements. 
    The gray dashed line denote the fit from extrapolated ECLAIRs points; the black line shows the fiducial broken power law fit to the EP-FXT/\emph{Swift}-XRT/\emph{Chandra} data; the magenta dash-dotted line shows the alternative Gaussian plus broken PL baseline fit; and the magenta dashed curve shows the smoothly broken power-law fit to the late X-ray component. Dashed purple and blue radio lines show the temporal fits for 5.5 and 7.4 GHz. \emph{Bottom}: Photon index evolution from the absorbed power-law fit as listed in Table~\ref{tab:xray_spec}.
    } 
  \label{fig:lightcurve}
\end{figure}

\section{DATA ANALYSIS}\label{sec:analysis}

\subsection{X-ray temporal and spectral evolution}

For a common representation of the X-ray evolution, the best-fit spectral models were used to derive unabsorbed flux densities at 1 keV for different instruments, as shown in Figure~\ref{fig:lightcurve}.
The temporal and spectral X-ray evolution can be described by $F_\nu\propto t^{-\alpha_\mathrm{x}}\nu^{-\beta_\mathrm{x}}$.
A double-broken power law fitted to the well-sampled EP-FXT and \emph{Swift}-XRT light curve gives three principal temporal segments: $\alpha_{\rm x,1} = 0.93^{+0.06}_{-0.06}$ before $t_{\rm x,1} = 1.27^{+0.02}_{-0.02} \times 10^4$ s, followed by a rapid decay with $\alpha_{\rm X,2}=5.28^{+0.11}_{-0.12}$, and a second shallow component after $t_{\rm X,2}=4.73^{+0.25}_{-0.24}\times10^4$ s with $\alpha_{\rm X,3}=0.42_{-0.07}^{+0.07}$.
The earlier ECLAIRs measurements were extrapolated to 1 keV using their best-fit spectral models, which suggest a near-flat evolution, with nominal estimated $\alpha_{\rm x,0}=0.03^{+0.06}_{-0.02}$ and $t_{\rm x,0}\simeq1620$ s (gray dashed line in Figure~\ref{fig:lightcurve}).

The X-ray emission before abrupt decay $t_{\rm x,1}$ constitutes a prolonged X-ray plateau, while the late shallow component emerging after $t_{\rm x,2}$ corresponds to the second plateau; and the two phases are analyzed separately below.
Unless otherwise stated, all uncertainties are quoted at the $1\sigma$ level.

\subsubsection{Early X-ray plateau and rapid decay}

We jointly fitted the EP-FXT and \emph{Swift}-XRT spectra with the \texttt{bayspec} package using an absorbed power law model, \texttt{tbabs*ztbabs*powerlaw}.
The Galactic column density was fixed at $N_{\rm H,Gal}=2.94\times10^{20}\ {\rm cm^{-2}}$ \citep{Willingale2013}. 
The joint fit of the high-count spectra before the end of rapid decay $t_{\rm x,2}$ gives an intrinsic column density at $z=0.505$ of $N_{\rm H,int}=4.0^{+0.2}_{-0.2}\times10^{21}\ {\rm cm^{-2}}$. 
The resulting time-resolved spectral parameters, unabsorbed fluxes, 1 keV flux densities and fit statistics are summarized in Table~\ref{tab:xray_spec}.

A pronounced hard-to-soft evolution was accompanied with the rapid decay, with the photon index evolving from $\Gamma_{\rm X}\simeq2.1$ to $\Gamma_{\rm X}\simeq3.6$.
Allowing $N_{\rm H,int}$ to vary independently during rapid decay, the fit improves only $\Delta\chi^2=4.4$ while $\Delta{\rm AIC}=+5.6$ with five additional parameters, and $\Gamma_{\rm x}$ still softens from $\simeq2.7$ to $\simeq3.5$
(Figure~\ref{fig:lightcurve}).
Thus, absorption evolution is not statistically required nor sufficient to explain the observed softening.
On the other hand, to assess the influence of the small flares superposed on the early X-ray component, we also fitted the light curve with Gaussian components plus an underlying broken power-law baseline, yielding $\alpha'_{\rm X,1}=0.86_{-0.05}^{+0.05}$ and $t'_{\rm X,1}=1.34_{-0.01}^{+0.01}\times10^4$ s, consistent with the description before.

\begin{figure}[t]
    \centering
    \begin{overpic}[width=0.45\textwidth]{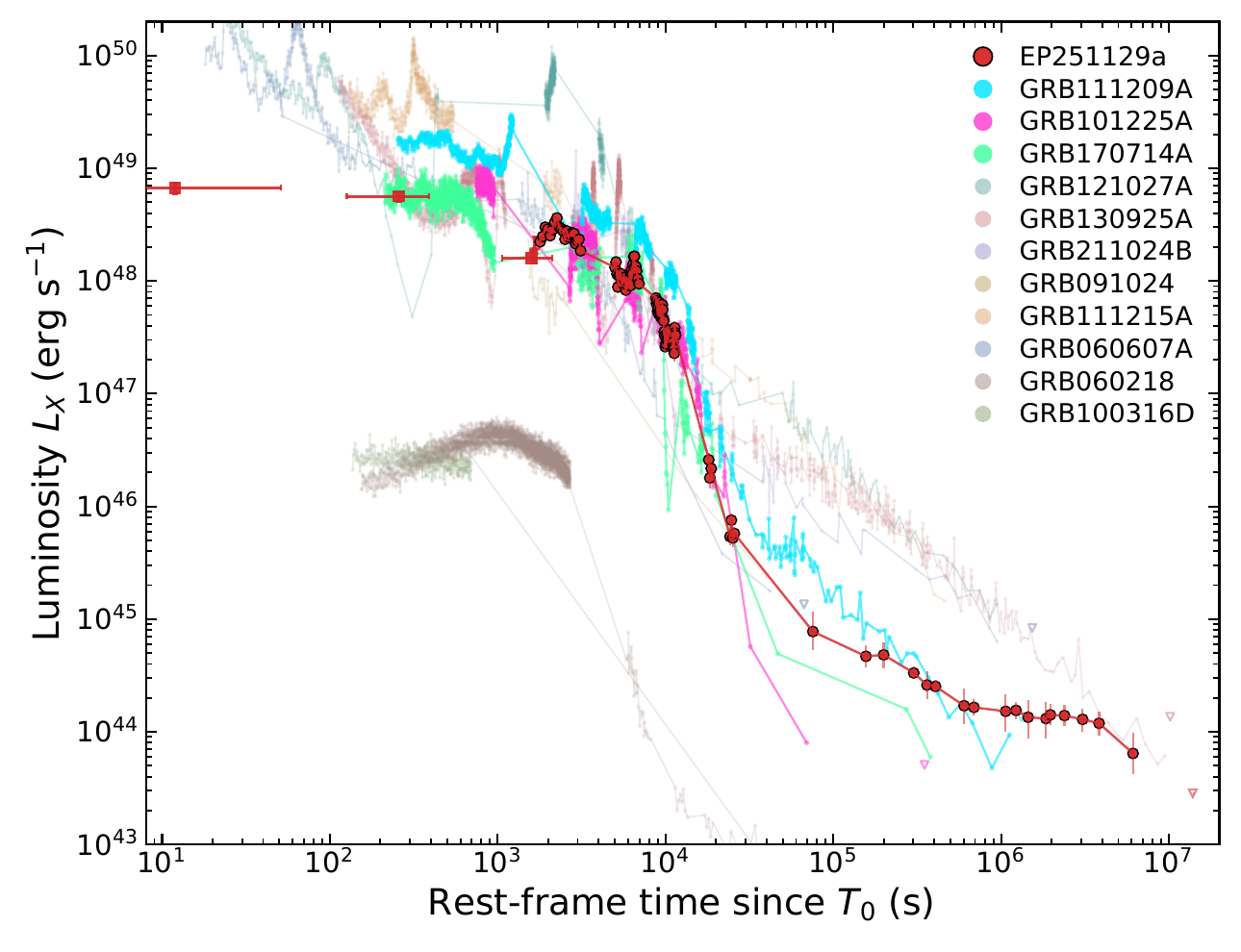}
    \end{overpic}
   \caption{ X-ray light curve of EP251129a compared with representative ULGRBs, low-luminosity GRBs (LLGRBs). All light curves have been converted to the 0.3 -- 10 keV band in the rest frame. }
\label{fig:comparaionX}
\end{figure}

The early X-ray emission exhibits a shallow decay accompanied by several small flares, and is terminated by an abrupt rapid decay, resembling the
prolonged internal plateaus of the ULGRBs, as shown in Figure~\ref{fig:comparaionX}. In particular, its early evolution is  similar to those of GRB~101225A, GRB~170714A, and less luminous than that of GRB~111209A.
In contrast, the early VT\_B and VT\_R light curves decline smoothly, with a single power law of slope about 0.7 over the first $\sim3000$ s, without counterparts to the X-ray flares or the abrupt X-ray decline \citep{GCN42887}, indicating that the early optical and X-ray emission are dominated by distinct components.

\subsubsection{Late-time X-ray plateau}

Since the limited photon statistics at late times, adjacent spectra were fitted jointly with tied photon indices, while the flux normalization was allowed to vary.
The late-time spectra give $N_{\rm H,int}=2.1^{+1.5}_{-1.3}\times10^{21}\ {\rm cm^{-2}}$.
Following the rapid decay, the spectrum hardens again to $\Gamma_{\rm X}\sim2$, while the light curve turns to an overall shallow decline with $\alpha_{\rm X,3}=0.42_{-0.07}^{+0.07}$. 
As a bandpass check, refitting the faint late-time EP-FXT spectra over 0.5--4 keV changes by less than 10\%, with the $1\sigma$ intervals overlapping those obtained over 0.5--10 keV.
To resolve the temporal substructure within the late component,  we also fitted the late light curve with a smoothly broken power law. 
The evolution contains an early bump with $\alpha'_{\rm x,1} = -0.12^{+0.42}_{-0.48}$ and $\alpha'_{\rm x,2}=1.12^{+0.57}_{-0.32}$,
separated at $t'_{\rm b,1}=3.1^{+1.4}_{-0.7}\times10^5$ s, followed by an extended shallow phase with $\alpha'_{\rm x,3}=0.08^{+0.17}_{-0.13}$. 
A subsequent steepening is inferred around $t'_{\rm b,2} = 5.4^{+1.8}_{-1.6}\times10^6$ s, with $\alpha'_{\rm x,4} = 1.45^{+1.83}_{-0.85}$, which is further constrained by the deep \emph{Chandra} upper limit at $\sim242$ days. 
This exceptionally long rest-frame break time $\sim42$ days places EP251129a at the extreme long-duration end of the GRB X-ray plateau population \citep{DongXF2026}.
The spectral hardening and long-lived shallow evolution after the rapid tail, suggests that the late X-ray plateau is dominated by an emerging component distinct from the early X-ray.

\subsection{Radio afterglow}

For the first five VLA C-band epochs, the 4--8 GHz data were divided
into four sub-bands to construct near-simultaneous radio spectra, while the final two broadband epochs were divided into 11 representative
sub-bands spanning 1.3--11.5 GHz (Table~\ref{tab:vla1}).
The in-band spectral indices show substantial variation in early epochs, which may be affected by calibration covariance and interstellar scintillation. %\citep{Rickett1990,Goodman1997}
For the temporal analysis and broadband modeling below, we use the 5.5 and 7.4 GHz measurements for representative.

\begin{figure*}[tbp]
    \centering
    % \vspace{0.2cm}
    % 第二行子图
    \begin{overpic}[width=0.364\textwidth]{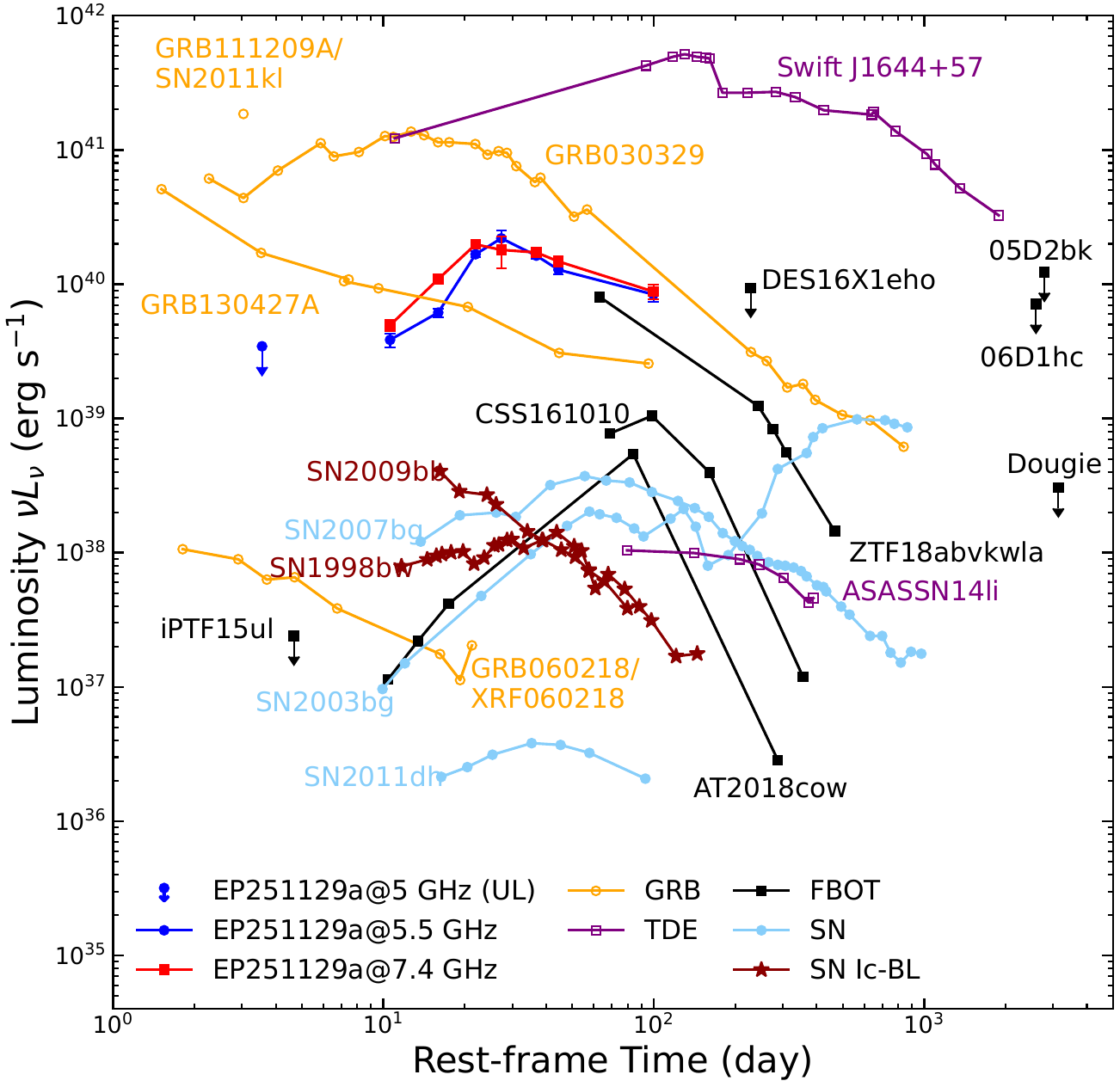}
        % \put(-1,102){\textbf{(a)}} % 标注
    \end{overpic}
    \hspace{0.05\textwidth} % 调整间距
    \begin{overpic}[width=0.370\textwidth]{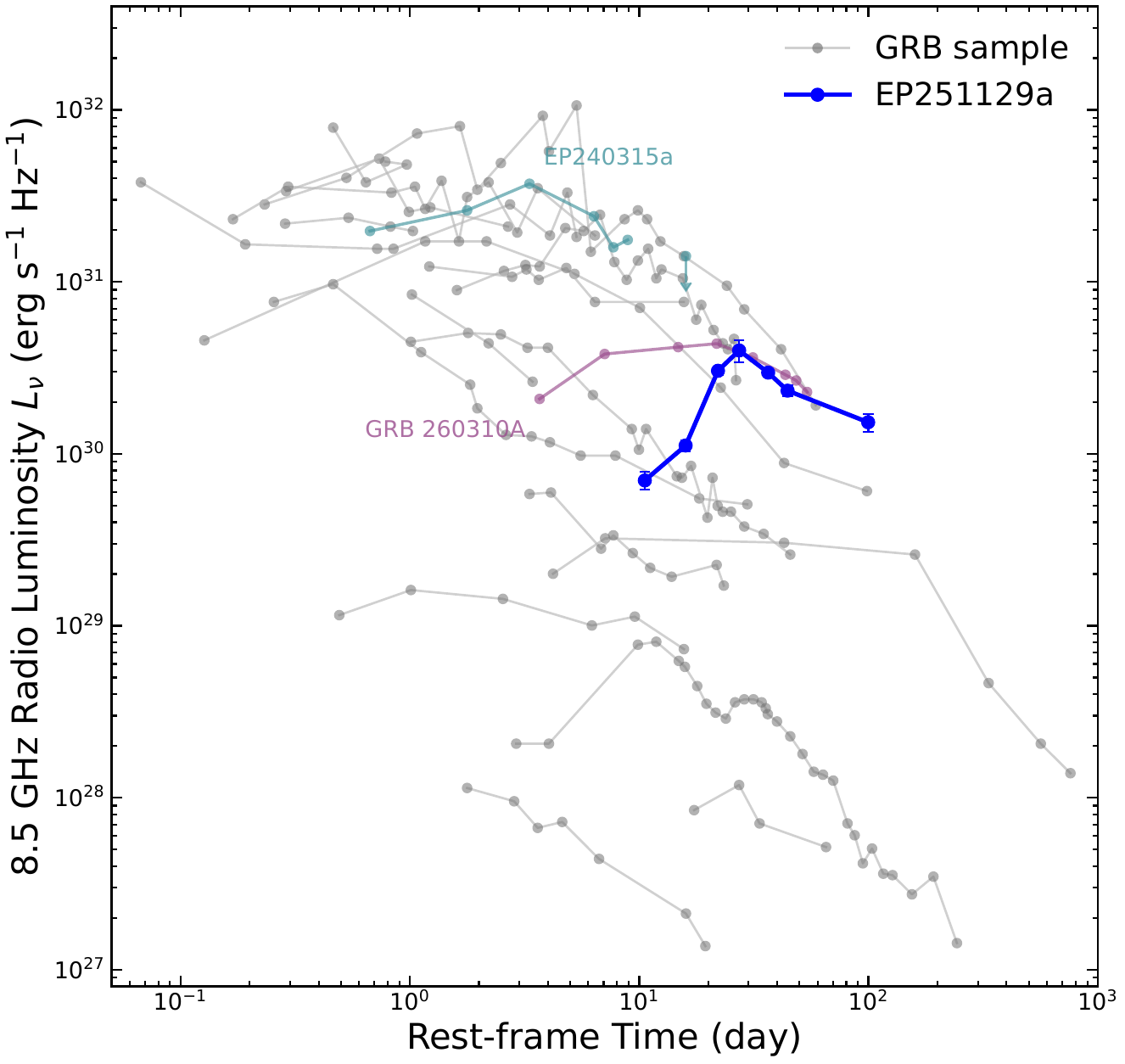}
        % \put(-1,99){\textbf{(b)}} % 标注
    \end{overpic}

    % 总体标题
   \caption{  \emph{Left}: The 5.5 GHz (blue) and 7.4 GHz (red) radio luminosity of EP251129a compared with low-frequency (1–10 GHz) light curves of different classes of energetic explosions: tidal disruption events, SNe, relativistic Ic-BL SNe, long-duration GRBs, and fast blue optical  transients. \emph{Right}: The interpolated 8.5 GHz Radio luminosity light curve of EP251129a (blue) compared with the sample of cosmological long GRBs and sub-energetic GRBs in the rest frame \citep{Chandra2012, Ricci2025, YangYH2026}.}
\label{fig:comparaion}
\end{figure*}

The radio emission remains faint initially and then rises rapidly, with the temporal indices  $\alpha_{\rm 5.5,1}=-2.46^{+0.25}_{-0.26}$ , $\alpha_{\rm 7.4,1}=-1.89_{-0.13}^{+0.13}$, and then decay with $\alpha_{\rm 5.5,2}=0.71^{+0.14}_{-0.13}$ and $\alpha_{\rm 7.4,2}=0.64^{+0.14}_{-0.13}$ after the peak. The 5.5 GHz and 7.4 GHz light curves peak at approximately 36--40 days, corresponding to $\sim24$--27 days in the rest frame. 
The peak flux density of $\simeq0.38$ mJy corresponds to $\nu L_\nu\simeq3\times10^{40}\ {\rm erg\,s^{-1}}$, within the luminosity range of GRB radio afterglows (Figure~\ref{fig:comparaion}a). 
However, its temporal evolution is unusual among long GRBs, whose radio afterglows peak mostly at rest-frame times of $\sim$3--6 days, substantially earlier than EP251129a, and generally exhibit a much shallower rise \citep{Chandra2012}.
Notably, the early non-detection of e-MERLIN at $\sim4.6$--6 days followed by the rapid rise in VLA and much delayed peak, indicates that the radio component emerges only at substantially later times.
The measured pre-peak rise is also steeper than expected from a post-deceleration on-axis forward shock. For $\nu_a<\nu<\nu_m$, the standard scalings are $F_\nu\propto t^{1/2}$ in a uniform ISM and $F_\nu\propto t^0$ in a wind environment, while even in a strong self-absorbed regime with $\nu_m<\nu<\nu_a$, the steepest rise scalings are $t^{5/4}$ and $t^{7/4}$, respectively \citep[e.g.,][]{Granot2002,Zhang2018Book}. The observed $t^{1.9-2.5}$ rise therefore exceeds these standard expectations. 

Together, the early radio upper limit, delayed peak, and rapid rising, indicate an off-axis geometry, in which the energetic jet becomes visible as the blast wave decelerates and the relativistic beaming cone widens toward the LOS
\citep{Gill2018,Pang2024a,Pang2024b}.
The contemporaneous shallow X-ray plateau further indicates additional energy injection into the blast wave, which could also modify the afterglow evolution \citep{ZhangB2006,Zhao2020}.
The off-axis geometry and late energy injection are modeled quantitatively in Section~\ref{sec:afterglow}.

\section{PHYSICAL INTERPRETATION AND MODELING}
\label{sec:model}
The observations above indicate that the early X-ray emission is distinct from the optical and later X-ray/radio components. 
The early X-ray plateau terminated by abrupt rapid decay is consistent with characteristics of the internal plateau, which is attributed to the internal dissipation of a long-lived central engine. In contrast, the smoothly declining early optical emission suggests an early afterglow component, while the second X-ray plateau and the rapidly rising, delayed radio afterglow motivate an off-axis jet core with additional energy injection.
We therefore propose a scenario in which the first X-ray plateau is powered by a newborn millisecond magnetar undergoing GW-dominated spin-down before collapsing into a BH, whereas the optical, radio, and late X-ray emission arise from an off-axis two-component forward shock with additional energy injection, possibly powered by fallback accretion onto the BH. 
Alternative interpretations are discussed and compared in Section~\ref{sec:con}.

\begin{figure*}[htbp]
    \centering
    % 第一行子图
    \begin{overpic}[width=0.37\textwidth]{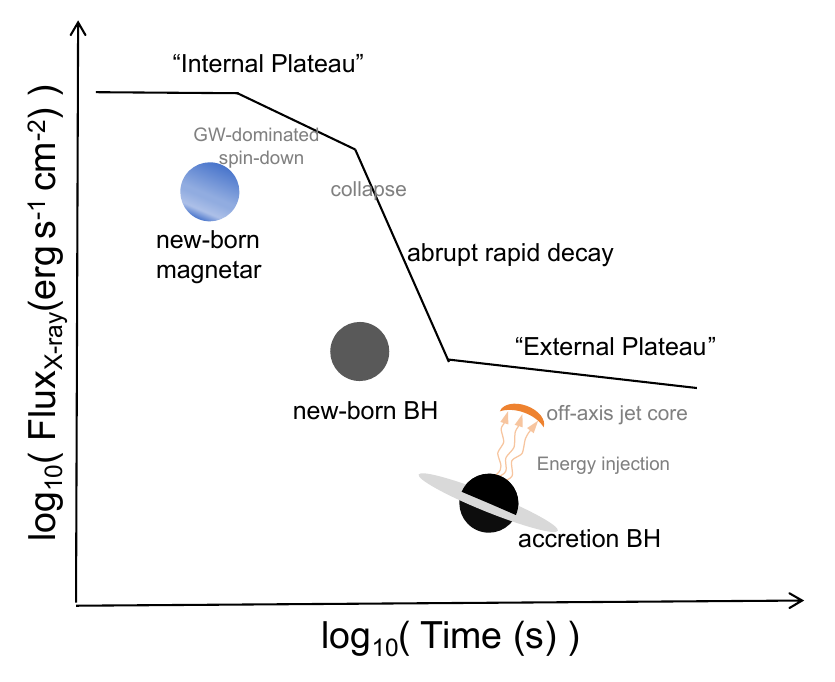}
    \end{overpic}
    \hspace{0.01\textwidth} % 调整间距
    \begin{overpic}[width=0.245\textwidth]{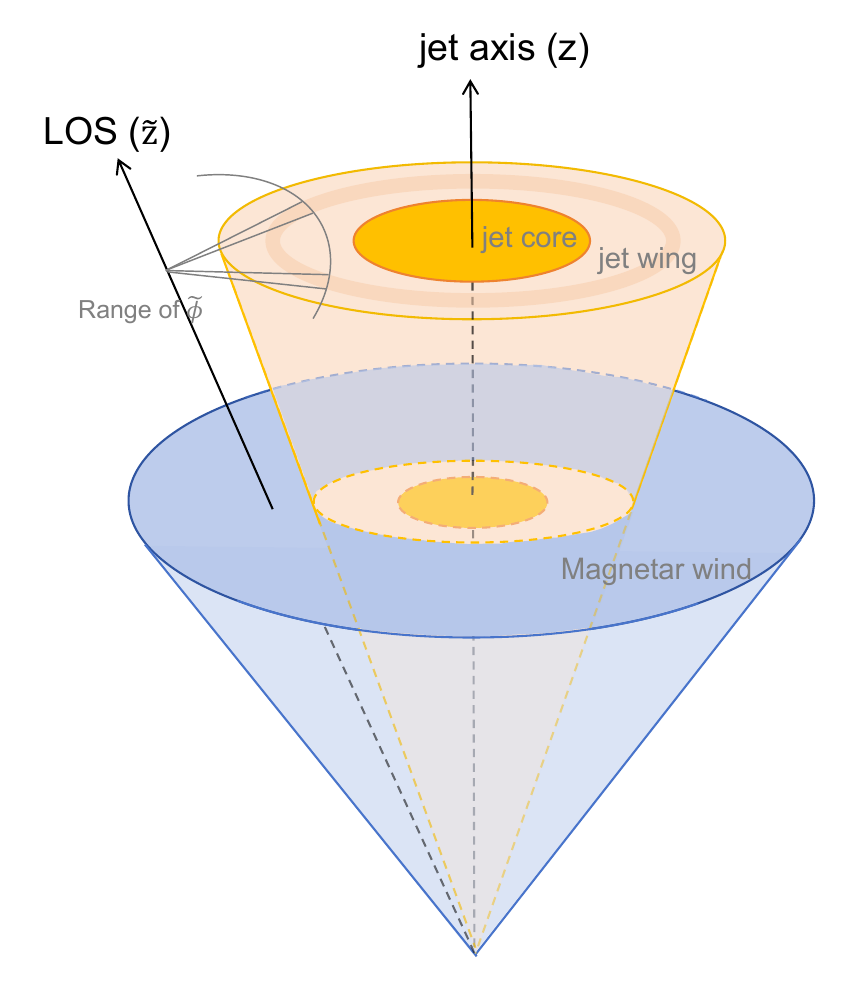}
    \end{overpic}
    
\caption{ \textbf{Schematic of the preferred model.} \emph{Left}:  The early internal X-ray plateau is powered by the dipole radiation wind of newborn-magnetar during GW-dominated spin-down, and the collapse to a BH produce the abrupt rapid decay. At later times, fallback accretion onto the newborn BH can power additional BZ energy injection into the off-axis afterglow.
\emph{Right}:  Approximate coaxial geometry of the wide magnetar wind (blue) and the off-axis two-component afterglow jet (orange), consisting of an energetic core surrounded by a less energetic wing. The jet direction is along the z direction, while the LOS is along the $\tilde{z}$ direction. The light orange shaded ring denotes an arbitrary segment of the jet. The sketch is not to scale.}
\label{fig:model_schematic}
\end{figure*}

\subsection{Early X-ray plateau and magnetar collapse} \label{subsec:internal_plateau}

In the newborn-magnetar scenario, the early X-rays are produced by internal dissipation of the relativistic magnetar wind \citep[e.g.][]{Xiao2017,Xiao2018,Lv2018}.
The rotational energy of the magnetar is lost through magnetic-dipole (MD) and
gravitational-wave (GW) quadrupole radiation, 
with the characteristic spin-down timescales $\tau_{\mathrm{c}, \mathrm{gw}}  \simeq 9.1 \times 10^3 \, \mathrm{s} \, I_{45}^{-1} \epsilon_{-3}^{-2} P_{0,-3}^4 $, and $\tau_{\mathrm{c}, \mathrm{md}}  \simeq 2.05 \times 10^3 \, \mathrm{s} \, I_{45} B_{\mathrm{p}, 15}^{-2} P_{0,-3}^2 R_6^{-6} $,  where $B_{\mathrm{p}, 15}$, $R_6$, $I_{45}$ and $\epsilon$ are the surface polar magnetic field, radius, moment of inertia and ellipticity of the magnetar \citep{Lv2018}.
For a newborn rapidly-rotating magnetar with significant ellipticity, the initial spin-down can be dominated by GW losses, for which the electromagnetic luminosity evolves as $L_{\rm EM}(t)\simeq L_{{\rm em},0}(1+{t}/{\tau_{c,{\rm gw}}})^{-1}$, where the initial electromagnetic luminosity $L_{\mathrm{em}, 0}  \simeq 1.0 \times 10^{49} B_{\mathrm{p}, 15}^2 P_{0,-3}^{-4} R_6^6 \ \mathrm{erg} \mathrm{~s}^{-1}$ \citep{Lv2018}. 
Assume the magnetar wind is collimated with half opening angle $\theta_\mathrm{mw}$ \citep{Bucciantini2008,Xie2020,Zou2021}, the observed X-ray luminosity can be related to the magnetar wind power as $L_\mathrm{X,iso}=\eta_\mathrm{x}L_\mathrm{EM}/f_\mathrm{b}$, where $\eta_{\rm X}$ is the X-ray radiation efficiency and $f_\mathrm{b}=1-\cos\theta_{\rm mw}$ is the beaming factor.
As the angular frequency decreases, the spin-down would transition to MD-dominant at $\tau_\star={\tau_{\mathrm{c}, \mathrm{md}}}/{\tau_{\mathrm{c}, \mathrm{gw}}}\left(\tau_{\mathrm{c}, \mathrm{md}}-2 \tau_{\mathrm{c}, \mathrm{gw}}\right)$.
And if continued spin-down reduces the rotational support below that required to sustain the supramassive magnetar, the magnetar would collapse into a BH, leading to the abrupt cessation of internal X-ray emission \citep{Troja2007,Lv2018}.

For EP251129a, the initial GW-dominant spin-down and collapse before transition to MD-dominant could naturally produce the observed early X-ray evolution.
Comparing with the X-ray luminosity $L_{\rm X,iso,0}\simeq4.3\times10^{48}\ {\rm erg\,s^{-1}}$ and the timescales $\tau_{\rm c,gw}\simeq t_{\rm x,0}/(1+z) \simeq 1.1 \times10^3$ s and $\tau_\ast\gtrsim t_{\rm x,1}/(1+z) \simeq 8.4\times10^3$ s, the constrain of $P_0$, $B_p$ and $\epsilon$ can be obtained as  $P_0\lesssim1.1(f_{\rm b}/\eta_{\rm X})^{-1/2}$ ms, $B_{\rm p}\lesssim7.3\times10^{14} (f_{\rm b}/\eta_{\rm X})^{-1/2}$ G, and $\epsilon\lesssim3.2\times10^{-3} (f_{\rm b}/\eta_{\rm X})^{-1}$.

Although the exact radiation physics of the magnetar wind remains uncertain, internal magnetic-dissipation models predict a curved spectrum rather than single power law, in which optical emission below synchrotron self-absorption frequency can be strongly suppressed, with the gamma-ray tail substantially weaker than the simultaneous X-ray emission \citep{Xiao2018,Xiao2017,Xiao2019}.
Therefore the smoothly declining early optical emission can be dominated by the afterglow, while the X-ray emission is accounted for the internal dissipation of magnetar wind.
In addition, the multiple X-ray flares superposed on the plateau are compatible with intermittent magnetic activity in a differentially rotating millisecond magnetar, powered by repeated magnetic-field amplification and reconnection \citep{Dai2006}.

\subsection{High-latitude effect and spectral softening}\label{sec:curvature}
\begin{figure*}[bp]

    \centering
    % 第一行子图 (a) 和 (b)
    \begin{overpic}[width=0.37\textwidth]{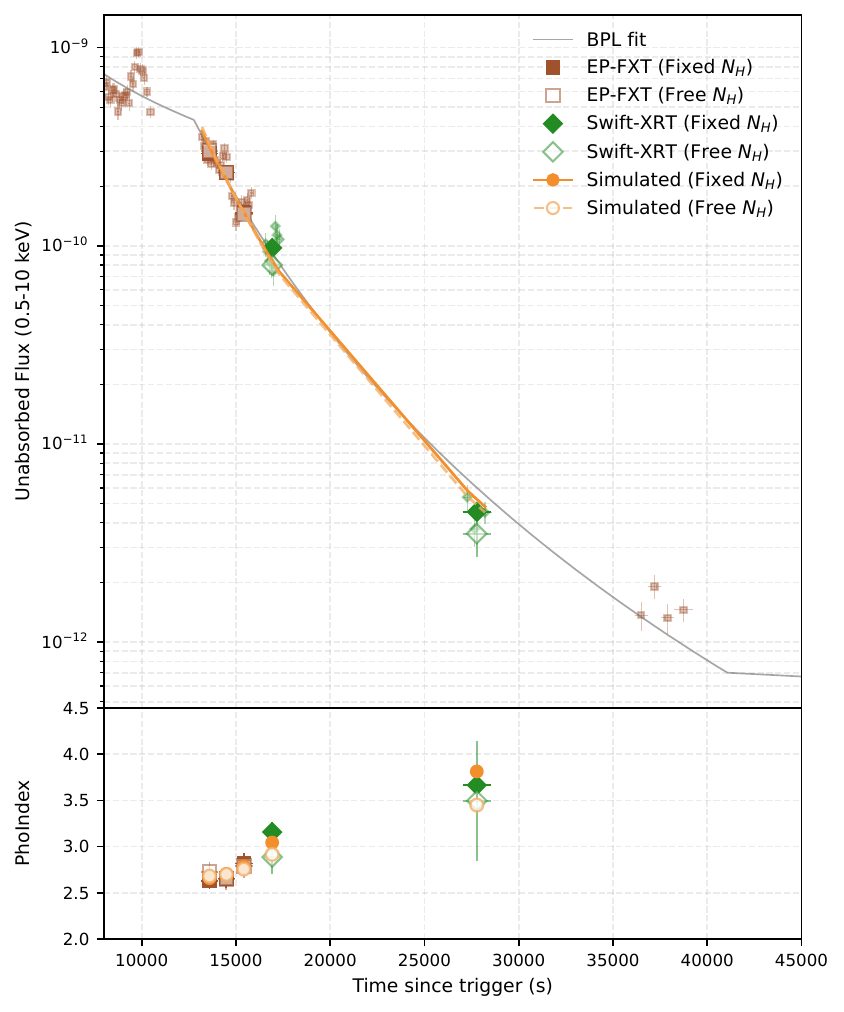}
    %     \put(2, 80){\textbf{(b)}} % 标注
    \end{overpic}
    \hspace{0.01\textwidth} % 调整间距
    \begin{overpic}[width=0.425\textwidth]{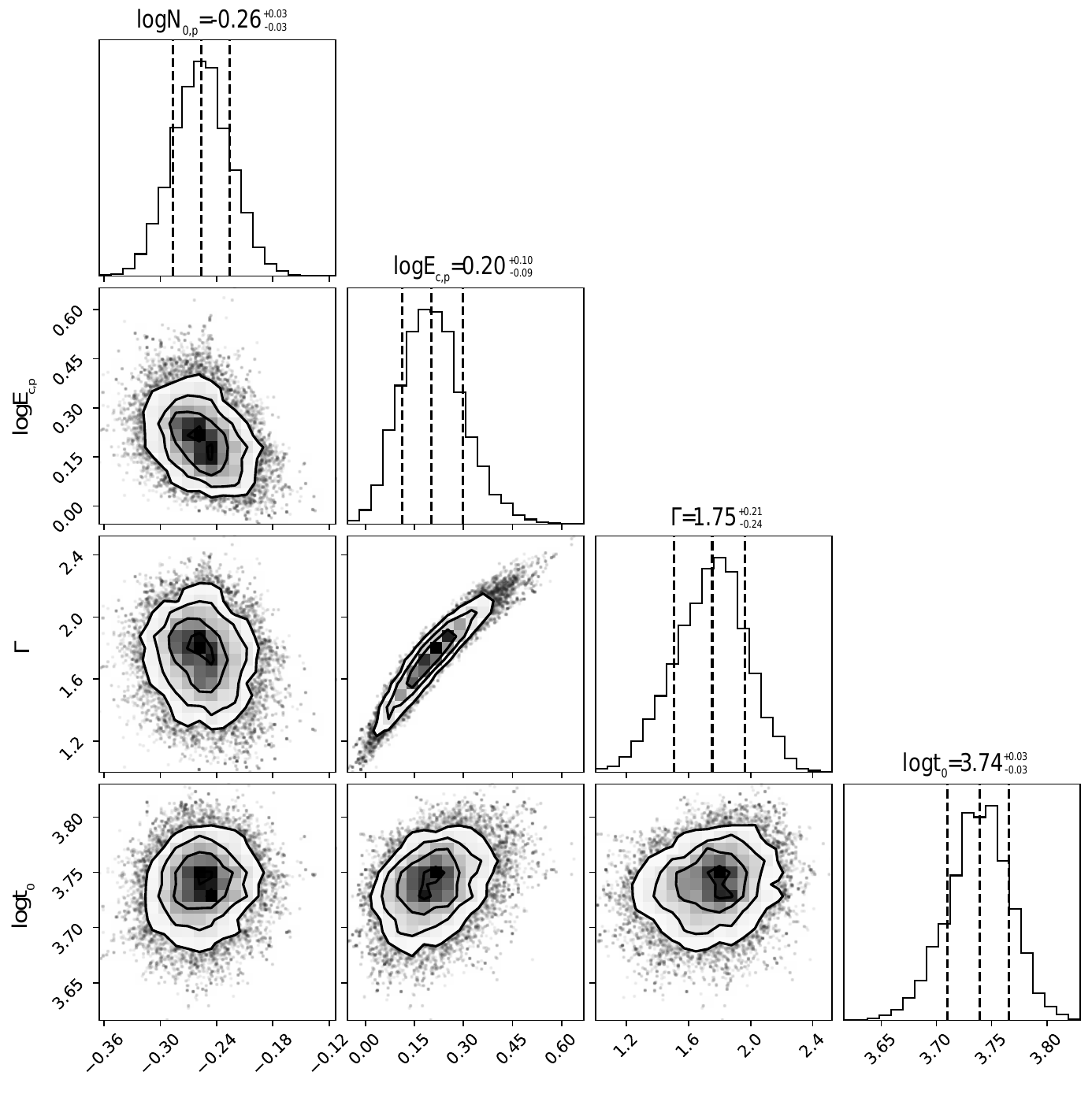}
        % \put(2,80){\textbf{(a)}} % 标注 a
    \end{overpic}
    % 总体标题
    \caption{\textbf{High latitude effect modeling of the rapid X-ray decay.} \emph{Left}: 0.5-10 keV X-ray lightcurve and photon index.
    Filled (open) symbols denote the spectral fits with fixed (freely varying) $N_{\rm H,int}$. 
    The orange symbols show the forward simulations of the evolving cutoff-power-law curvature model, while the brown and green symbols represent the time-average spectral fitting results of the EP-FXT and \emph{Swift}-XRT data. 
    \emph{Right}: MCMC fitting results for the time-dependent cutoff power-law (CPL) model in Section \ref{sec:curvature}. }
    \label{fig:curvature}
\end{figure*}

The cessation of the central engine can be followed by a high-latitude tail. % following the cessation of the central engine.
For a simple instantaneous power-law spectrum, the expected temporal decay is $F_\nu\propto(t-t_0)^{-(2+\beta_{\rm x})}$, where $t_0$ is the effective zero time of relevant emission episode for a long-lived central engine, rather than the GRB trigger time \citep{Kumar2000,Liang2006}.
The rapid decay of EP251129a is accompanied by strong spectral softening, which cannot be fully described by the high-latitude tail with constant spectrum.
We instead adopt a more general model of \citet{Zhang2009}, in which the instantaneous spectrum is described by an evolving cutoff power law,
\begin{equation}\label{eq_NEt}
N(E,t)=N_0(t)\left(\frac{E}{1\ {\rm keV}}\right)^{-\Gamma_{\rm ph}}
\exp\!\left[-\frac{E}{E_c(t)}\right],
\end{equation}
where the evolving cutoff energy $E_\mathrm{c}(t)=E_\mathrm{c,p} \left[ \frac{t-t_0}{t_\mathrm{p}-t_0}\right]^{-1}$, the spectral normalization $N_0(t)=N_\mathrm{0,p} \left[ \frac{t-t_0}{t_\mathrm{p}-t_0} \right]^{-(1+\Gamma_{\rm ph})}$ and $t_p$ is taken as the onset time of the rapid decay.
Since independent CPL fits within the narrow 0.5--10 keV band are strongly degenerate, we follow the ``forward simulation" as \cite{Zhang2009}: the evolving modeled spectra are folded through the corresponding instrumental responses and compared with the observed fluxes and effective photon indices.
As shown in Figure \ref{fig:curvature}, this model can reproduce both the rapid flux decay and spectral softening, %for either fixed or freely varying intrinsic absorption.
with $E_{\rm c,p}=1.58^{+0.41}_{-0.30}$ keV, $\Gamma_{\rm ph}=1.75^{+0.21}_{-0.24}$, and an effective zero time
$t_0=(5.5\pm0.4)\times10^3$ s.
Considering the typical emission timescale $\Delta t\sim (1+z)\frac{R_\mathrm{w}}{2c\Gamma_\mathrm{w}^2} \sim t_p-t_0$, and the maximum high-latitude delay $t_{\rm HLE,max}\gtrsim (1+z)\frac{R_\mathrm{w}\Delta\theta^2}{2c}\sim t_\mathrm{x,2}-t_\mathrm{x,1}$, where $\Delta\theta$ is the angle between the emitting edge and the LOS, the extent of the magnetar wind can be constrained as $\Delta\theta\gtrsim 1.2^{\circ} \Gamma_\mathrm{w,2}^{-1}$. 
%The duration of the curvature tail further requires an angular extent of the emitting magnetar wind of $\Delta\theta\gtrsim1.2^\circ\Gamma_{\rm w,2}^{-1}$ around the LOS.
For the coaxial wind--jet geometry illustrated in Figure~\ref{fig:model_schematic}, this implies that the magnetar wind is substantially wider than the narrow afterglow jet and encompasses the LOS, with the wind half-opening angle 
$\theta_{\rm mw}\gtrsim\theta_{\rm obs}+\Delta\theta\sim16^\circ$

\subsection{Off-axis afterglow and late energy injection}
\label{sec:afterglow}

Motivated by the discussion in Section \ref{sec:analysis}, we model the optical, radio, and late X-ray emission as synchrotron radiation from an off-axis two-component forward shock with additional energy injection.
The jet consists of a narrow energetic core with half-opening angle $\theta_{\rm c}$ and isotropic-equivalent kinetic energy $E_{\rm c}$, surrounded by a wider, less energetic wing extending to $\theta_{\rm j}$ with energy $E_{\rm w}$.
The circumburst density is described by $\rho=AR^{-2}$, with $A=5\times10^{11}A_\ast~{\rm g\,cm^{-1}}$ as expected for a massive-star wind \citep{Chevalier2000}.
The blast-wave dynamics, synchrotron radiation including self-absorption, and equal-arrival-time-surface integration  for an off-axis observer follow the established treatments of \citet{Huang1998,Huang1999,Zou2005,Zhang2018Book,Pang2024a,Pang2024b}.
Following the collapse of the magnetar, a fraction of the gravitationally bound stellar material can fall back onto the newborn BH, and the energy of the BH could be extracted and eventually injected into the blast wave through the Blandford-Znajek (BZ) mechanism \citep{Zhao2020,Chen2017}.
The fallback rate after the formation of the BH is parameterized as
\citep{Wu2013,Zhao2020}
\begin{equation}
\dot M_{\rm fb}\! =\! 2\dot M_p\! \left[ \left(\frac{t-t_{\rm inj,0}}{t_{\rm inj,p}\!-t_{\rm inj,0}}\right)^{\!-\frac{1}{2} }
\!\!\!+\!
\left(\frac{t-t_{\rm inj,0}}{t_{\rm inj,p}\!-t_{\rm inj,0}}\right)^{\!\frac{5}{3} } \right]^{-1}
\end{equation}
for $t>t_{\rm inj,0}$, where $t_{\rm inj,0}$ and $t_{\rm inj,p}$ are the start and peak time in the local frame, $ \dot M_{\rm p}$ is the peak fallback rate. The mass accretion rate onto the BH is smoothed by the disk viscosity \citep{Kumar2008a,Zhao2020},
\begin{equation}
\frac{d\dot M}{dt} = \frac{\dot M_{\rm fb}-\dot M}{\tau_{\rm vis}}.
\end{equation}
The BH accretion rate is smoothed over the disk viscous timescale $\tau_{\rm vis}$, so that the injected BZ luminosity does not directly track the instantaneous fallback rate. Especially for a large viscous timescale, the BH accretion rate can remain flat until $t > \tau_\mathrm{vis}$ and then decrease as the fallback rate.
The Blandford--Znajek luminosity is calculated following 
\citet{Chen2017,Zhao2020}, 
\begin{equation}
L_{\rm BZ}\simeq
9.3\times10^{53}
\frac{a_\bullet^2 F(a_\bullet)}
{\left(1+\sqrt{1-a_\bullet^2}\right)^2}\!
\left(\!\frac{\dot M}{M_\odot~{\rm s}^{-1}}\!\right)
{\rm erg~s^{-1}},
\end{equation}
where $F(a_\bullet)=[(1+q^2)/q^2][(q+1/q)\arctan q -1]$ and $q=a_\bullet/(1+\sqrt{1-a_\bullet^2})$. We adopt the initial mass and spin of the BH $M_{\bullet,0}= 2.2\,\mathrm{M_\odot}$ and $a_{\bullet,0}=0.4$ as fiducial values, and we assume the Poynting-dominated injected energy is deposited primarily into the energetic jet core. 
% The fallback rate is characterized by the start time $t_{\rm inj,0}$, peak time $t_{\rm inj,p}$ in the local frame, and peak fallback rate $\dot M_{\rm p}$.
%The detailed prescriptions are summarized in Appendix~\ref{supp_afterglow}.

\begin{figure}[tbp]
    \centering
    % 第一行子图
    \hspace{0.0\textwidth} % 调整间距
    \begin{overpic}[width=0.45\textwidth]{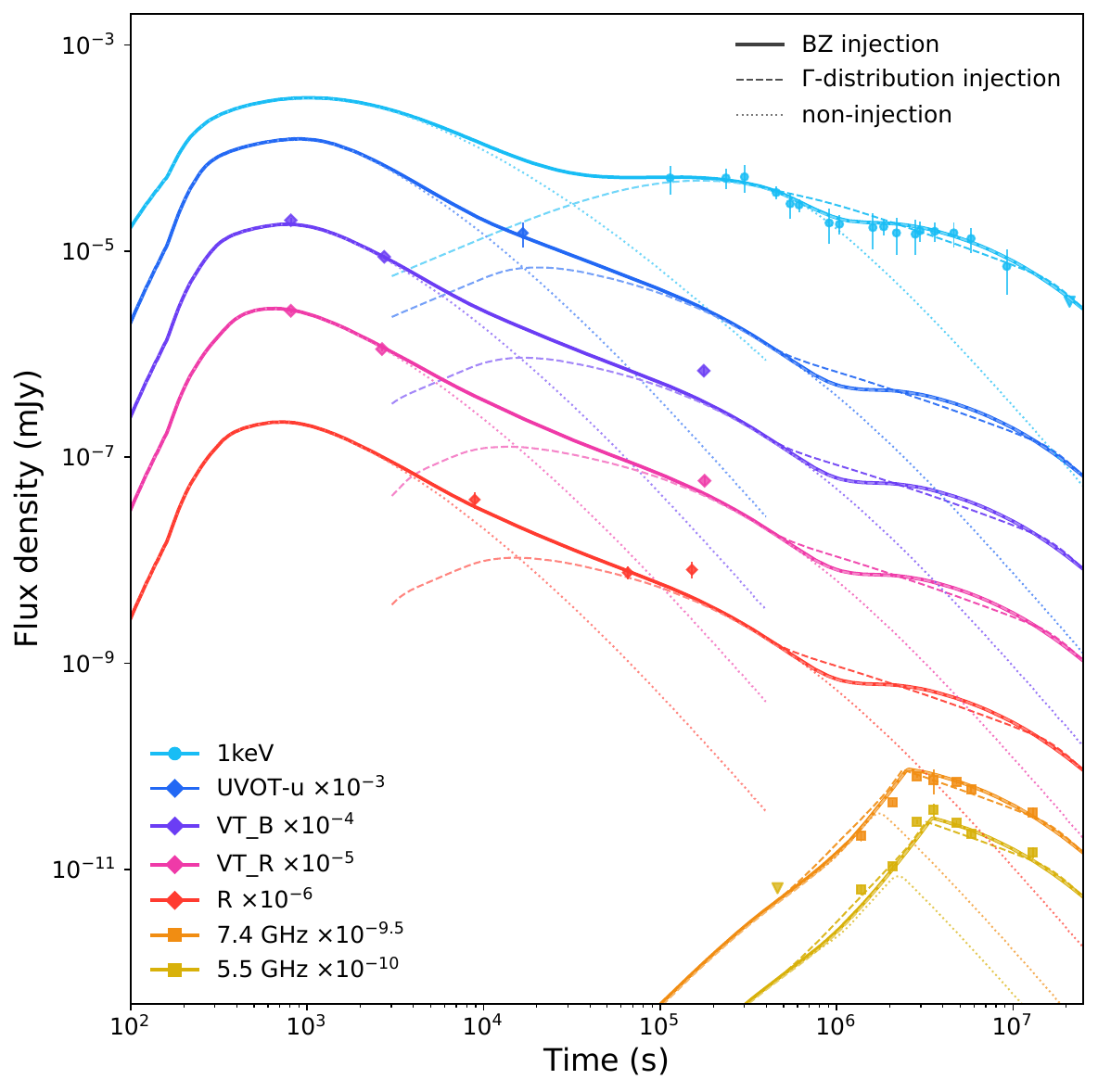}
    \end{overpic}
    % 总体标题
    \caption{\textbf{Multi-wavelength afterglow modeling of EP251129a.} The early optical, late X-ray, and radio light curves are modeled with an off-axis two-component forward shock in a wind environment.
    The solid, dashed, and dotted curves show the BZ- injection, Lorentz-factor-stratified, and no-injection models, respectively. 
    The early internal X-ray plateau and its curvature tail are excluded from the afterglow fit.
    The best-fit parameters are $E_{\rm c} = 2.4 \times 10^{53} \, \rm erg$, $E_{\rm w} = 3.9 \times 10^{51} \, \rm erg$, $\theta_{\rm obs} =14.6^{\circ} $, $\theta_{\rm c} = 3.9^{\circ}$,  $\theta_{\rm j} =6.5^{\circ} $, $p =2.12 $, $\epsilon_e = 0.32 $, $\epsilon_B = 6.2\times 10^{-6}$, $A_* =4.6 $,  $t_{\rm inj,p} = 3.2 \times 10^{5} \, \rm s$, $\tau_{\rm vis} = 6.7\times 10^{5} \, \rm s$  and $\dot{M}_{\rm inj,p} = 1.7\times 10^{-7} \, M_\odot \, \rm s^{-1}$. The distributions of the parameters are shown in Figure~\ref{fig_af_contour} and  Table~\ref{tab_afterglow_mcmc_results}. 
    }
    \label{fig:afterglow_fit}
\end{figure}

We fit the early optical, radio, and late X-ray with MCMC method, excluding the early X-ray plateau and its curvature tail. 
The initial Lorentz factors are fixed at $\Gamma_{0,\rm c}=\Gamma_{0,\rm w}=200$, since the limited early optical coverage does not strongly constrain their initial values and the later evolution is insensitive to this choice. The start time of fallback accretion is also less constrained, and we fixed at $t_{\rm inj,0} = 5.0\times 10^{4} \, \rm s$. 
The posterior constraints are summarized in Table~\ref{tab_afterglow_mcmc_results}, and the best-fit light curves are shown in Figure~\ref{fig:afterglow_fit}.
The model can well reproduce the broadband afterglow observations. The early optical emission is dominated by the wing closer to the LOS, whereas the late X-ray and radio emission arise from the energetic core, as the blast wave decelerates and its beaming cone widens toward the LOS.
The total BZ-injected energy $E_{\rm inj}\simeq6\times10^{51}$ erg  corresponds to an accreted mass onto BH of $M_{\rm acc}\simeq0.2\,M_\odot$.
Without additional energy injection, the model underpredicts the radio maximum and produces a substantially steeper late X-ray decline.

As an alternative to continued central-engine activity, we also implement a Lorentz-factor-stratified  shock model, in which the ejecta have a Lorentz-factor distribution $M(>\Gamma)\propto\Gamma^{-s}$.
Slower ejecta catch up with the decelerating blast wave and gradually increase its energy \citep{Rees1998,Sari2000,ZhangB2006}.
The reduced $\chi^2$ are 1.1, 3.1, and 58, for the BZ-injection, $\Gamma$-stratified, and no-injection models, respectively; and $\Delta{\rm AIC}=43$ and 1469 of the latter two relative to the BZ-injection model.
Therefore, the BZ-injection model is regarded as the preferred description for the afterglow.
The underpredicted late optical emission $\gtrsim10^5\,\rm s$ suggest an additional component, possibly cocoon emission, and is discussed in Section~\ref{sec:con}.

\begin{table}[t]
\centering
\scriptsize
%\begin{threeparttable}
\caption{MCMC parameter estimate (median) for the afterglow model with 1$\sigma$ uncertainties.}\label{tab_afterglow_mcmc_results}
\begin{tabular}{lcccc}
\toprule
Parameter & Unit & Prior Type & Prior & Posterior \\ 
% & & (Power-law jet) & (Gaussian jet)  \\
\hline
$\theta_{\rm obs}$ & deg & uniform    & [0, 25]   &  $12.6^{+5.1}_{-2.3}$ \\
$\theta_{\rm c}$   & deg & uniform  & [0, $\theta_{\rm j}$ ] & $3.4^{+2.8}_{-1.1}$ \\
$E_{\rm c}$        & erg &log-uniform & [50, 55]    & $53.48^{+0.23}_{-0.40}$  \\
$\Gamma_{\rm c}$   & 1   & Fixed      & 200                      & --   \\
$\theta_{\rm j}$   & deg & uniform    & [0, 25]   & $ 6.3^{+4.5}_{-1.7} $  \\
$E_{\rm w}$        & erg &log-uniform & [49, $\log E_{\rm c}$ ]  &  $51.39^{+0.17}_{-0.23}$  \\
$\Gamma_{\rm w}$   & 1   & Fixed      & 200                      & --   \\
$p$                & 1   &uniform     & [2.01,3]    & $ 2.13^{+0.01}_{-0.02}$  \\
$\epsilon_{\rm e}$ & 1   &log-uniform & [-2, -0.22]  &  $ -0.54^{+0.16}_{-0.10}$   \\
$\epsilon_{\rm B}$ & 1   &log-uniform & [-7, -1]    & $-5.00^{+0.21}_{-0.25}$  \\
$A_\star$          & 1   &log-uniform & [-1,2]     & $0.62^{+0.09}_{-0.08}$ \\
% \hline
$t_{\rm inj,0}$    & s   & Fixed  &  $5\times10^4\rm\,s$    &  --\\
$t_{\rm inj,p}$    & s   &log-uniform &  [$\log t_{\rm inj,0}$ , 7]  &  $5.54^{+0.08}_{-0.11}$ \\
$\dot{M}_{\rm inj,p}$& $M_\odot\,{\rm s^{-1}}$ &log-uniform & [-8, -5]  & $-6.87^{+0.23}_{-0.21}$ \\
$\tau_{\rm vis}$    & s   &log-uniform &  [$\log t_{\rm inj,0}$ , 7]  & $5.75^{+0.15}_{-0.11}$ \\
\bottomrule
\end{tabular}
%\end{threeparttable}
\end{table}

\section{Discussion and Conclusions} \label{sec:con}
EP251129a was active in the ECLAIRs 4--120 keV band for more than $3\times10^3$ s, with a relatively hard initial spectrum ($\Gamma\simeq1.7$) that subsequently softened and connected smoothly to the EP X-ray activity extending to $\sim10^4$ s.
This resembles key identifying properties of ULGRB, whose high-energy emission lasting $\sim10^4$ s within two orders of magnitude of the peak, followed by a rapid decay with $\alpha>3$, and often accompanied by variability during the prolonged activity \citep{Levan2014,Fu2024}.
Together with the late optical rebrightening consistent with an associated supernova, these properties favor an X-ray-rich SN-associated ULGRB from a massive-star explosion.

Although the magnetar scenario provides a natural interpretation for the early internal X-ray plateau with fewest assumptions, a BH central engine could in principle also power the early long-lived X-ray emission through fallback accretion and the BZ mechanism \citep{Wu2013,Gao2016}.
For an on-axis observer, the isotropic-equivalent X-ray luminosity can be connected to the BZ power as $L_\mathrm{X,iso}^{\mathrm{on}} = \eta_\mathrm{X}L_\mathrm{BZ}/f_b$, where $\eta_\mathrm{X}$ and $f_b$ are the radiation efficiency and beaming factor \citep{Kumar2008a,Lei2013,Wu2013,Gao2016}.
The observed $t^{-1}$ evolution would require specific progenitor density and angular-momentum distributions, or a long viscous timescale to regulate the BH accretion rate \citep{Kumar2008a,Lei2017,Zhao2020}.
However, the early accretion-powered X-ray emission must terminate abruptly at $t_{\rm x,1,rest}$, which is difficult to reconcile with the smooth viscous modulation \citep{Kumar2008a}, while a second more extended phase of mass supply is required at later times. 
The two temporally separated episodes would require additional assumptions about the radial structure of the fallback material or a rapid evolution efficiency, beyond a simple smooth fallback prescription.
The off-axis geometry inferred from the late afterglow introduces an additional constraint. For an early BH jet sharing the same narrow geometry as the late afterglow, the observed internal X-ray emission would be strongly Doppler deboosted. The off-axis specific flux is related to its on-axis emission by, $F_{\nu,\mathrm{off}}(\Delta\theta,t) = a^3F_{\nu/a,\mathrm{on}}(0,at) = a^{3+\beta_X}F_{\nu,\mathrm{on}}(0,at)$ for the assumed local power-law spectrum $F_\nu\propto\nu^{-\beta_X}$, and $a=\mathcal D_{\rm off}/\mathcal D_{\rm on}\simeq[1+\Gamma_X^2\Delta\theta^2]^{-1}$, $\Delta\theta=\theta_\mathrm{obs}-\theta{\mathrm{j}}$ \citep{Granot2002b,Yamazaki2002}.
For the best-fit afterglow geometry and $\Gamma\sim200$, the Doppler ratio $a\sim10^{-3}$, implying severe suppression of the observed X-ray emission. Reproducing the bright early X-ray component would require either an extremely large intrinsic luminosity or a substantially wider jet cone encompassing the line of sight.
Thus, a BH interpretation is not excluded, but requires additional temporal and geometrical assumptions.

For the late afterglow, a Lorentz-factor-stratified model provides an alternative to continued central-engine activity \citep{Rees1998,Sari2000,ZhangB2006}.
However, its gradual energy transfer with relatively smooth dynamical evolution cannot reproduce the late X-ray turnovers and is statistically disfavored by the comparison in Section~\ref{sec:afterglow}. By contrast, the Poynting-dominated BZ injection can add substantial energy with little additional mass and produce a stronger change in the late evolution.
Discrete late-shell collisions could also re-energize the blast
wave, but energetic collisions would generally produce stronger
flares or multi-band rebrightenings than observed here \citep{Geng2025,Fu2026}.
Therefore fallback-powered BZ injection is regarded as the preferred model.

The forward-shock model does not reproduce all of the optical
emission, particularly the blueward excess at $\sim10^5$ s. A simple blackbody estimate with $T_{\rm BB}\sim10^4$ K and $L_{\rm bol}\sim5\times10^{43}\ {\rm erg\,s^{-1}}$, is consistent with a luminous cocoon-cooling emission with extended envelope or circumstellar material \citep{Hamidani2025a, Hamidani2025b, ZhengJH2026}.
Detailed early radio spectral turnovers are also more complex than the smooth afterglow model and may be affected by interstellar scintillation \citep{Rickett1990,Goodman1997}.

\vspace{0.8em}

In summary, EP251129a presents a distinctive combination of an ULGRB-like internal X-ray plateau followed by an abrupt spectral-softening rapid decay, an independently evolving optical component, an exceptionally long second X-ray plateau, and an unusual rapid-rising radio afterglow with a delayed peak.
These observations allow the long-lived central-engine emission and the off-axis external-shock afterglow to be disentangled within a single event, and are self-consistently described by a newborn magnetar undergoing GW-dominated spin-down and collapse into a BH, followed by fallback-powered energy injection into an off-axis two-component jet.
EP251129a therefore connects the emerging EFXT population with ULGRB-like central-engine activity and off-axis afterglow, offering a rare opportunity to probe both central-engine evolution and jet geometry within a single event.

\begin{acknowledgments}
This work is based on the data obtained with \emph{Einstein Probe}, a space mission supported by the Strategic Priority Program on Space Science of Chinese Academy of Sciences, in collaboration with the European Space Agency, the Max-Planck-Institute for extraterrestrial Physics (Germany), and the Centre National d'Études Spatiales (France). This work made use of data supplied by the UK Swift Science Data Centre at University of Leicester.
We thank Jiapei Feng, Yi-Han Yin, Jun Yang, Jingwei Hu, Dongyue Li, Feifan Song, Yijia Zhang, Qinyu Wu, Chen Wang, Wei Chen, Guoli Huang for arranging EP observations and discussions.
Y. W. thank Dekun Song for helpful suggestions and discussions.
This work was supported by the National Natural Science Foundation of China (grant No. 12393812) and the Strategic
Priority Research Program of the Chinese Academy of Sciences (grant No. XDB0550300). 
T. A. acknowledges support from the Shanghai Oriental Talent Project. 
A.L. W. acknowledges support from the Specialized Research Fund for State Key Laboratory of Radio Astronomy and Technology Project.
We would like to thank the referee for his/her helpful comments that allowed us to improve our manuscript significantly.

\end{acknowledgments}

\facilities{
\emph{Einstein Probe} (WXT, FXT),
Space Variable Objects Monitor (ECLAIRs, VT),
\emph{Swift} (XRT, UVOT),
\emph{Fermi} (GBM),
\emph{Chandra} (ACIS),
VLA,
e-MERLIN,
GTC, OSN, SAO RAS and Mondy
}

%%%%%%%%%%%%%%%%%Appendix%%%%%%%%%%%%%%%%%%%%%%%%

% \clearpage
% \appendix
% \vspace{-0.8em}

%%%%%%%%%%%%%%%%%%%Figures/Tables%%%%%%%%%%%%%%%%%%%%%%%%%%%%%

\begin{figure*}[bp]

    \centering
    % 第一行子图 (a) 和 (b)
    \begin{overpic}[width=0.47\textwidth]{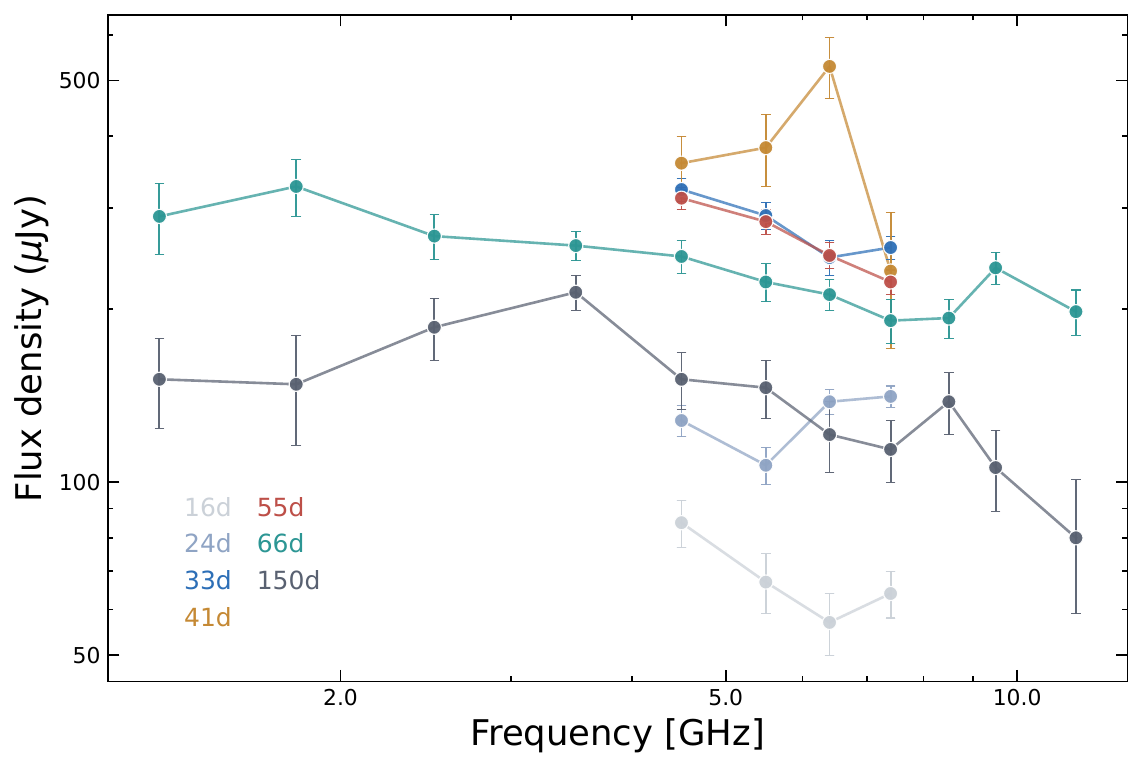}
    %     \put(2, 80){\textbf{(b)}} % 标注
    \end{overpic}
    \hspace{0.01\textwidth} % 调整间距
    \begin{overpic}[width=0.32\textwidth]{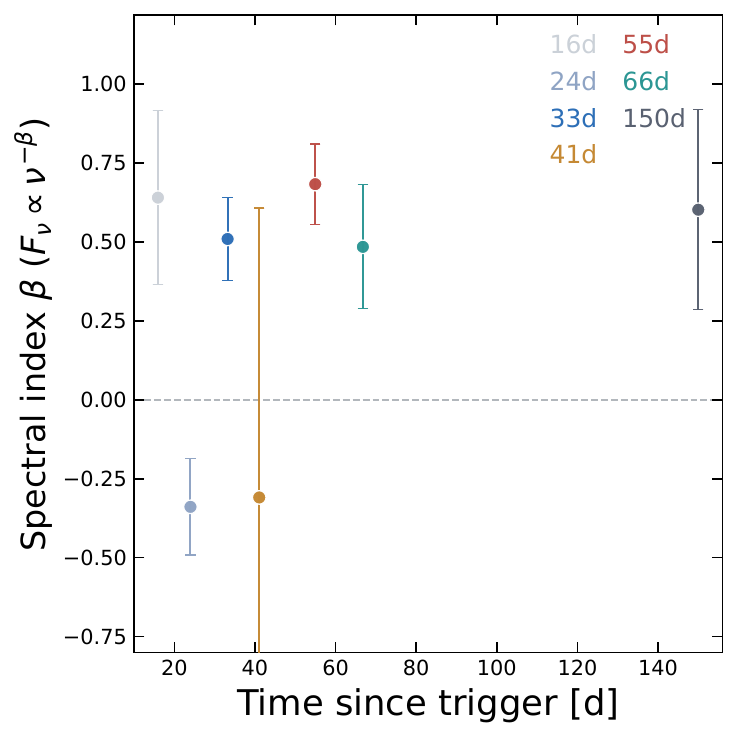}
        % \put(2,80){\textbf{(a)}} % 标注 a
    \end{overpic}
    % 总体标题
    \caption{\emph{Left Panel}: Radio spectral energy distributions (SEDs) of EP251129a of seven epochs. \emph{Right Panel}: The 4.5 - 7.4 GHz spectral index evolution. }
    \label{fig:Radio_SED}
\end{figure*}

% \clearpage

\begin{figure*}[htbp]
    \centering
    % 第一行子图
    \begin{overpic}[width=1.0\textwidth]{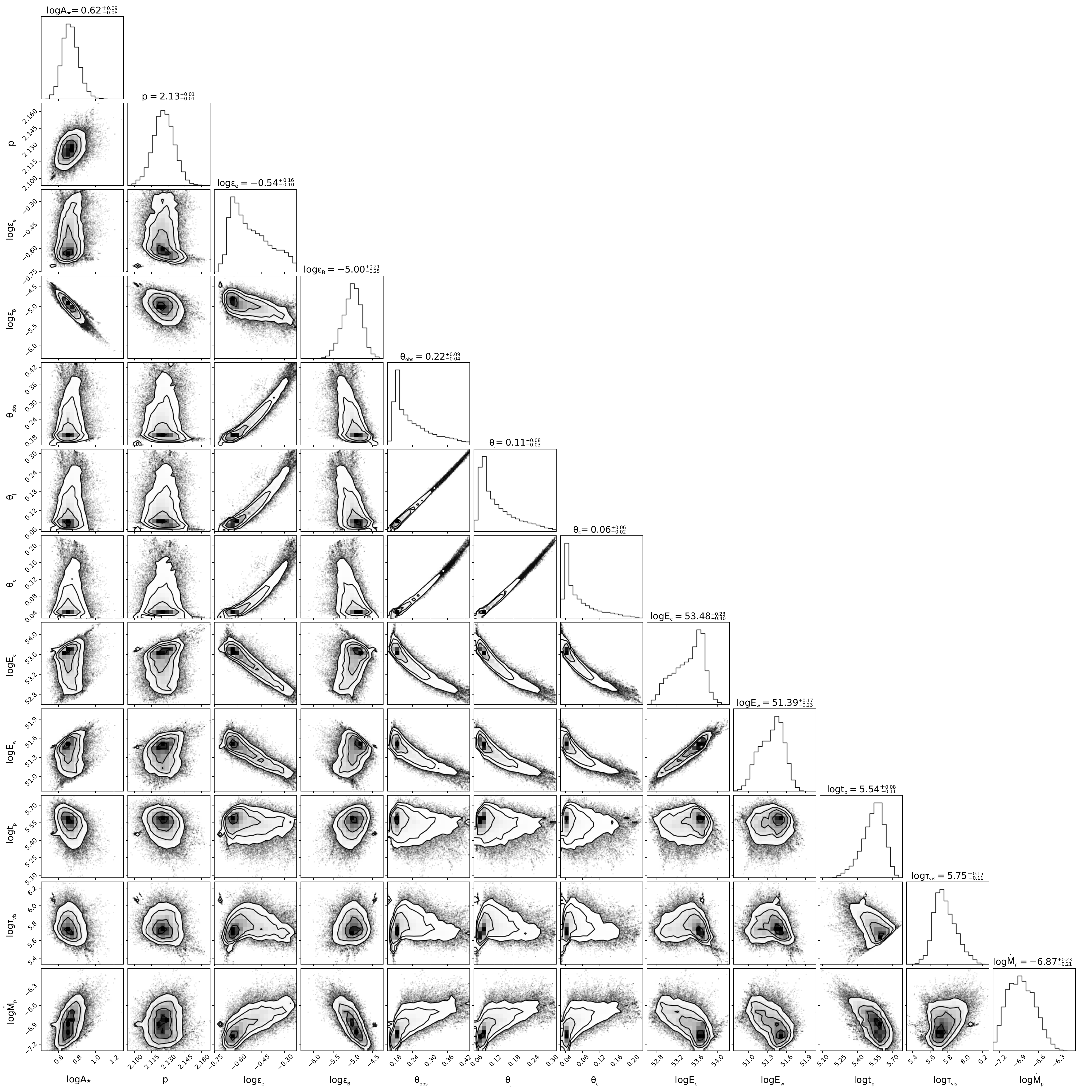}
        %\put(2,92){\textbf{(a)}} % 标注
    \end{overpic}

    \caption{\textbf{MCMC fitting results for multi-band afterglow modeling.} The afterglow emission is modeled as the synchrotron emission from a forward shock expanding in a wind environment, assuming an off-axis two-component jet with late-time energy injection (see S\ref{sec:afterglow} for details). The diagonal dashed lines and labels indicate the median values along with the 1$\sigma$ uncertainties for the marginalized distribution of each parameter.}

    \label{fig_af_contour}
\end{figure*}

\begin{table*}
\centering
\scriptsize
\begin{threeparttable}
\small
\caption{  Log of the X-ray observations of EP251129a.} 
\label{tab_FXT_XRT_obs}
\begin{tabular}{ccccc}
\toprule
ObsID & Start$\;$Time & End$\;$Time & Duration & Exposure\\
 & (UTC) & (UTC) & (second) & (second)\\

\midrule
\multicolumn{5}{c}{\textbf{EP-WXT}}\\
\midrule
11916652534 & 2025-11-29T03:09:02 & 2025-11-29T03:14:24 & 322 & 82 \\
 
\midrule
\multicolumn{5}{c}{\textbf{EP-FXT}}\\
\midrule
01709248891 & 2025-11-29T03:16:08 & 2025-11-29T05:25:08 & 7740 & 5004.9 \\
06800001018 & 2025-11-29T05:26:09 & 2025-11-29T06:56:52 & 5443 & 2786.3 \\
08500000454 & 2025-11-29T12:34:30 & 2025-11-29T13:26:02 & 3092 & 3091.8 \\
06800001029 & 2025-12-04T07:30:51 & 2025-12-04T09:58:48 & 8877 & 9387.8 \\
06800001038 & 2025-12-06T02:36:27 & 2025-12-06T06:40:29 & 14642 & 9449.1 \\
06800001046 & 2025-12-11T00:46:39 & 2025-12-11T04:45:51 & 14352 & 8564.1 \\
08500000486 & 2025-12-20T09:58:20 & 2025-12-20T14:03:26 & 14706 & 9592.6 \\
06800001095 & 2026-01-02T14:12:34 & 2026-01-02T17:21:31 & 11337 & 5974.4 \\
06800001105 & 2026-01-09T14:06:19 & 2026-01-09T18:49:05 & 16966 & 8925.7 \\
06800001124 & 2026-01-18T19:13:31 & 2026-01-18T23:23:15 & 14984 & 9198.1 \\
06800001132 & 2026-01-23T23:48:11 & 2026-01-24T04:05:27 & 15436 & 7149.3 \\
06800001148 & 2026-02-03T21:41:07 & 2026-02-04T02:04:53 & 15826 & 8459.5 \\
06800001268 & 2026-03-15T14:01:32 & 2026-03-15T18:22:37 & 15665 & 8852.3 \\

\midrule
\multicolumn{5}{c}{\textbf{\emph{Swift}-XRT}}\\
\midrule
00021892001 & 2025-11-29T07:04:54 & 2025-11-29T07:21:06 & 972 & 842.5 \\
00021892002 & 2025-11-29T09:59:40 & 2025-11-29T10:27:28 & 1668 & 1494.4 \\
00021892003 & 2025-11-30T09:10:06 & 2025-11-30T11:05:53 & 6947 & 2685.3 \\
00021892004 & 2025-12-01T17:33:59 & 2025-12-01T22:29:55 & 17756 & 5636.4 \\
00021892005 & 2025-12-02T10:43:57 & 2025-12-02T17:14:23 & 23426 & 2269.1 \\
00021892006 & 2025-12-05T00:28:57 & 2025-12-05T19:24:22 & 68125 & 4412.8 \\
00021892007 & 2025-12-09T09:31:56 & 2025-12-09T19:13:21 & 34885 & 4295.0 \\
00021892009 & 2025-12-17T09:01:57 & 2025-12-17T23:16:20 & 51263 & 3557.9 \\
00021892010 & 2025-12-24T01:38:57 & 2025-12-24T15:52:52 & 51235 & 4187.2 \\
00021892011 & 2025-12-31T01:57:57 & 2025-12-31T16:08:20 & 51023 & 5408.2 \\

\midrule
\multicolumn{5}{c}{\textbf{\emph{Chandra}}}\\
\midrule
32430 & 2026-07-28T21:51:52 & 2026-07-29T01:21:41  & 12589 & 9944.8 \\

\bottomrule
\end{tabular}
\end{threeparttable}
\end{table*}

\clearpage
\begin{table*}
\centering
\setlength{\tabcolsep}{3.5pt}
\renewcommand{\arraystretch}{1.35}
\footnotesize
\caption{
Spectral results of EP-FXT and \emph{Swift}-XRT observations of EP251129a. Errors represent the 1$\sigma$ uncertainties.}
\label{tab:xray_spec}
\begin{tabular*}{\textwidth}{@{\extracolsep{\fill}}llccccc}
\hline
Time interval &  Mid time  &  Instrument & $\Gamma_X$ & Flux$^{\ddagger}$ & Flux density at 1 keV &  stat/dof  \\
&  &   &  & (${\rm erg,cm^{-2},s^{-1}}$) & (Jy) &  \\
\hline
$[2644,4749],{\rm s}$ & 3697 s & EP-FXT & $2.13^{+0.03}_{-0.03}$ & $1.98^{+0.03}_{-0.03}\times10^{-9}$ & $3.01^{+0.07}_{-0.06}\times10^{-4}$ &  131.23/149 \\
$[7486,10383],{\rm s}$ & 8935 s &  EP-FXT & $2.49^{+0.04}_{-0.04}$ & $6.54^{+0.12}_{-0.13}\times10^{-10}$ & $1.22^{+0.03}_{-0.03}\times10^{-4}$ & 108.06/90 \\
$[10445,15887],{\rm s}$ & 13166 s & EP-FXT & $2.65^{+0.04}_{-0.04}$ & $2.35^{+0.04}_{-0.04}\times10^{-10}$ & $4.69^{+0.11}_{-0.11}\times10^{-5}$ & 109.96/118 \\
$[16499,17341],{\rm s}$ & 16920 s & \emph{Swift}-XRT & $3.15^{+0.11}_{-0.10}$ & $9.78^{+0.49}_{-0.51}\times10^{-11}$ & $2.16^{+0.11}_{-0.11}\times10^{-5}$ & 30.07/37 \\
$[27029,28523],{\rm s}$ & 27776 s & \emph{Swift}-XRT & $3.64^{+0.21}_{-0.17}$ & $4.55^{+0.41}_{-0.39}\times10^{-12}$ & $9.88^{+0.90}_{-0.80}\times10^{-7}$ & 18.04/13 \\
$[36146,39237],{\rm s}$ & 37692 s & EP-FXT & $3.44^{+0.13}_{-0.14}$ & $1.51^{+0.10}_{-0.09}\times10^{-12}$ & $3.35^{+0.23}_{-0.21}\times10^{-7}$ & 61.25/68 \\
\hline

$[1.28,1.36],{\rm d}$ & 1.32 d & \emph{Swift}-XRT & $3.30^{+0.46}_{-0.56}$ & $2.30^{+0.81}_{-0.65}\times10^{-13}$ & $4.85^{+1.75}_{-1.37}\times10^{-8}$ & 3.78/6 \\

$[2.63,2.83],{\rm d}$ & 2.73 d & \emph{Swift}-XRT & $2.34^{+0.42}_{-0.38}$ & $2.94^{+0.71}_{-0.59}\times10^{-13}$ & $4.90^{+1.39}_{-1.08}\times10^{-8}$ & 4.96/8 \\

$[3.34,3.61],{\rm d}$ & 3.48 d & \emph{Swift}-XRT & $2.00^{+0.43}_{-0.32}$ & $3.82^{+1.29}_{-0.99}\times10^{-13}$ & $5.23^{+1.69}_{-1.34}\times10^{-8}$ & 9.17/6  \\

\hline

$[5.21,5.31],{\rm d}$ & 5.26 d & EP-FXT & \multirow{3}{*}{$1.92^{+0.12}_{-0.12}$} & $2.85^{+0.43}_{-0.38}\times10^{-13}$ & $3.70^{+0.46}_{-0.46}\times10^{-8}$ & \multirow{3}{*}{57.73/59}   \\
$[5.92,6.70],{\rm d}$ & 6.31 d  &  \emph{Swift}-XRT &  & $2.22^{+0.71}_{-0.57}\times10^{-13}$ & $2.82^{+0.99}_{-0.76}\times10^{-8}$ &  \\
$[7.00,7.17],{\rm d}$ & 7.09 d & EP-FXT &  & $2.17^{+0.34}_{-0.29}\times10^{-13}$ & $2.79^{+0.38}_{-0.32}\times10^{-8}$ &  \\
\hline

$[10.29,10.70],{\rm d}$ & 10.50 d & \emph{Swift}-XRT & \multirow{2}{*}{$1.97^{+0.30}_{-0.26}$} & $1.39^{+0.61}_{-0.44}\times10^{-13}$ & $1.89^{+0.78}_{-0.62}\times10^{-8}$ &  \multirow{2}{*}{10.50/15} \\
$[11.93,12.09],{\rm d}$ & 12.01 d & EP-FXT &  & $1.35^{+0.32}_{-0.26}\times10^{-13}$ & $1.81^{+0.37}_{-0.31}\times10^{-8}$ &  \\
\hline

$[18.27,18.86],{\rm d}$ & 18.57 d & \emph{Swift}-XRT & \multirow{6}{*}{$2.34^{+0.21}_{-0.22}$} & $9.74^{+4.18}_{-3.30}\times10^{-14}$ & $1.70^{+0.73}_{-0.57}\times10^{-8}$ &  \multirow{6}{*}{39.50/32} \\
$[21.31,21.48],{\rm d}$ & 21.40 d & EP-FXT &  & $9.92^{+1.83}_{-1.57}\times10^{-14}$ & $1.73^{+0.32}_{-0.27}\times10^{-8}$ &  \\
$[24.97,25.56],{\rm d}$ & 25.27 d & \emph{Swift}-XRT &  & $8.63^{+3.69}_{-3.02}\times10^{-14}$ & $1.50^{+0.64}_{-0.53}\times10^{-8}$ &  \\
$[31.98,32.57],{\rm d}$ & 32.28 d & \emph{Swift}-XRT &  & $8.40^{+3.45}_{-2.83}\times10^{-14}$ & $1.46^{+0.60}_{-0.49}\times10^{-8}$ &  \\
$[33.95,34.62],{\rm d}$ & 34.29 d & EP-FXT &  & $9.06^{+2.24}_{-1.75}\times10^{-14}$ & $1.58^{+0.39}_{-0.30}\times10^{-8}$ &  \\
$[41.48,41.68],{\rm d}$ & 41.58 d & EP-FXT &  & $8.93^{+2.12}_{-1.76}\times10^{-14}$ & $1.56^{+0.37}_{-0.31}\times10^{-8}$ &  \\
\hline

$[50.67,56.06],{\rm d}$ & 53.37 d &  EP-FXT & \multirow{3}{*}{$2.21^{+0.26}_{-0.30}$} & $9.34^{+2.66}_{-2.42}\times10^{-14}$ & $1.50^{+0.43}_{-0.36}\times10^{-8}$ &  \multirow{3}{*}{27.15/22} \\
$[66.78,66.98],{\rm d}$ & 66.88 d & EP-FXT &  & $8.24^{+2.46}_{-2.01}\times10^{-14}$ & $1.32^{+0.33}_{-0.29}\times10^{-8}$ &  \\
$[106.46,106.66],{\rm d}$ & 106.56 d &  EP-FXT &  & $4.42^{+2.54}_{-1.61}\times10^{-14}$ & $7.07^{+3.82}_{-2.45}\times10^{-9}$ &  \\
\hline
\end{tabular*}

\vspace{2mm}
\parbox{\textwidth}{\footnotesize
$^{*}$ All spectra are fitted with an absorbed power-law (PL) model in the $[0.5,10.0],{\rm keV}$ energy range.\\
$^{\dagger}$ The intrinsic absorption column at the redshift $z=0.505$ was constrained through a joint fit of the early spectra within $t_\mathrm{x,2}\simeq5\times10^4{\,\rm s}$, with $N_{\rm H,int}=4.0^{+0.2}_{-0.2}\times10^{21},{\rm cm^{-2}}$, and was constrained from the late spectra with $N_{\rm H,int} = 2.1_{-1.3}^{+1.5}\times10^{21}\ {\rm cm^{-2}}$.\\
$^{\ddagger}$ Since the limited photon statistics at late stages, spectra of adjacent time epochs were fitted jointly with tied photon indices.
}
\end{table*}

\begingroup
\scriptsize
\setlength{\tabcolsep}{4.5pt}
\renewcommand{\arraystretch}{0.90}
\setlength{\LTpre}{0pt}
\setlength{\LTpost}{0pt}
\setlength{\LTleft}{\fill}
\setlength{\LTright}{\fill}

\begin{longtable}{@{}lcccccc@{}}
\caption{VLA observations of GRB.}
\label{tab:vla1} \\
\toprule
Date$^{\rm a}$ & $\nu\,^{\rm b}$ & BMAJ$^{\rm c}$ & BMIN$^{\rm c}$ & BPA$^{\rm c}$ & RMS$^{\rm d}$ & Peak flux density$^{\rm e}$ \\
     & (GHz) & ('') & ('') & ($^{\circ}$) & ($\mu$Jy\,beam$^{-1}$) & ($\mu$Jy\,beam$^{-1}$) \\
\midrule
\endfirsthead

\multicolumn{7}{c}{\tablename~\thetable{} -- \textit{continued}} \\
\toprule
Date & $\nu$ & BMAJ & BMIN & BPA & RMS & Peak flux \\
     & (GHz) & ('') & ('') & ($^{\circ}$) & ($\mu$Jy\,beam$^{-1}$) & ($\mu$Jy\,beam$^{-1}$) \\
\midrule
\endhead

\midrule
\multicolumn{7}{r}{\textit{Continued on next page}} \\
\endfoot

\bottomrule
\noalign{\vskip 2.5pt}
\multicolumn{7}{@{}l@{}}{%
\begin{minipage}{0.90\textwidth}
\scriptsize
\setlength{\parskip}{0pt}
\setlength{\baselineskip}{9pt}
\raggedright
$^{\rm a}$ Observation dates are given in UT.\par
$^{\rm b}$ The observing frequency $\nu$ is the representative central frequency of each sub-band.\par
$^{\rm c}$ BMAJ, BMIN, and BPA denote the major axis, minor axis, and position angle of the synthesized beam, respectively; the beam position angle is measured east of north.\par
$^{\rm d}$ The RMS is the local image noise measured around the transient position.\par
$^{\rm e}$ Peak flux densities are measured at the VLA position of GRB 251129a in the image plane, assuming the source is unresolved. Quoted uncertainties are $1\sigma$ statistical errors.
\end{minipage}
} \\
\endlastfoot

2025-12-15 & 4.5 & 2.7 & 1.2 & 51 & 7.5 & $85 \pm 8$ \\
2025-12-15 & 5.5 & 2.3 & 0.97 & 51 & 7.7 & $67 \pm 8$ \\
2025-12-15 & 6.4 & 1.9 & 0.84 & 51 & 7.3 & $57 \pm 7$ \\
2025-12-15 & 7.4 & 1.7 & 0.75 & 52 & 5.6 & $64 \pm 6$ \\
\midrule
2025-12-23 & 4.5 & 3.1 & 1.1 & 20 & 7.6 & $128 \pm 8$ \\
2025-12-23 & 5.5 & 2.6 & 0.94 & 20 & 7.6 & $107 \pm 8$ \\
2025-12-23 & 6.4 & 2.2 & 0.82 & 20 & 7.2 & $138 \pm 7$ \\
2025-12-23 & 7.4 & 1.9 & 0.72 & 21 & 5.6 & $141 \pm 6$ \\
\midrule
2026-01-01 & 4.5 & 2.7 & 1.1 & $-$58 & 15 & $323 \pm 15$ \\
2026-01-01 & 5.5 & 2.2 & 0.94 & $-$58 & 16 & $291 \pm 16$ \\
2026-01-01 & 6.4 & 2 & 0.81 & $-$59 & 17 & $246 \pm 17$ \\
2026-01-01 & 7.4 & 1.7 & 0.72 & $-$57 & 12 & $256 \pm 12$ \\
\midrule
2026-01-09 & 4.5 & 3.3 & 1.1 & $-$15 & 41 & $359 \pm 41$ \\
2026-01-09 & 5.5 & 2.7 & 0.92 & $-$14 & 55 & $382 \pm 55$ \\
2026-01-09 & 6.4 & 2.3 & 0.78 & $-$16 & 64 & $529 \pm 64$ \\
2026-01-09 & 7.4 & 2 & 0.7 & $-$14 & 62 & $233 \pm 62$ \\
\midrule
2026-01-23 & 4.5 & 1.3 & 1.1 & 45 & 14 & $312 \pm 14$ \\
2026-01-23 & 5.5 & 1.1 & 0.94 & 43 & 14 & $284 \pm 14$ \\
2026-01-23 & 6.4 & 0.94 & 0.81 & 52 & 13 & $248 \pm 13$ \\
2026-01-23 & 7.4 & 0.84 & 0.7 & 46 & 11 & $223 \pm 11$ \\
\midrule
2026-02-03 & 1.3 & 4.3 & 3.5 & $-$12 & 41 & $290 \pm 41$ \\
2026-02-03 & 1.8 & 3.1 & 2.5 & $-$15 & 37 & $327 \pm 37$ \\
2026-02-03 & 2.5 & 2.1 & 1.6 & $-$10 & 24 & $268 \pm 24$ \\
2026-02-03 & 3.5 & 1.6 & 1.2 & $-$10 & 15 & $258 \pm 15$ \\
2026-02-03 & 4.5 & 1.3 & 0.96 & $-$7 & 16 & $247 \pm 16$ \\
2026-02-03 & 5.5 & 1.1 & 0.8 & $-$7 & 17 & $223 \pm 17$ \\
2026-02-03 & 6.4 & 0.82 & 0.6 & $-$11 & 13 & $212 \pm 13$ \\
2026-02-03 & 7.4 & 0.92 & 0.67 & $-$6 & 17 & $191 \pm 17$ \\
2026-02-03 & 8.5 & 0.67 & 0.5 & $-$2 & 15 & $193 \pm 15$ \\
2026-02-03 & 9.5 & 0.6 & 0.44 & $-$2 & 15 & $236 \pm 15$ \\
2026-02-03 & 11.5 & 0.51 & 0.37 & $-$9 & 18 & $198 \pm 18$ \\
\midrule
2026-04-28 & 1.3 & 3.1 & 1.3 & $-$38 & 27 & $151 \pm 27$ \\
2026-04-28 & 1.8 & 2.2 & 0.88 & $-$38 & 32 & $148 \pm 32$ \\
2026-04-28 & 2.5 & 1.6 & 0.61 & $-$32 & 23 & $186 \pm 23$ \\
2026-04-28 & 3.5 & 1.2 & 0.46 & $-$32 & 15 & $214 \pm 15$ \\
2026-04-28 & 4.5 & 0.94 & 0.36 & $-$29 & 17 & $151 \pm 17$ \\
2026-04-28 & 5.5 & 0.76 & 0.29 & $-$28 & 17 & $146 \pm 17$ \\
2026-04-28 & 6.4 & 0.67 & 0.26 & $-$29 & 17 & $121 \pm 17$ \\
2026-04-28 & 7.4 & 0.58 & 0.23 & $-$27 & 14 & $114 \pm 14$ \\
2026-04-28 & 8.5 & 0.54 & 0.2 & $-$19 & 17 & $138 \pm 17$ \\
2026-04-28 & 9.5 & 0.48 & 0.18 & $-$18 & 17 & $106 \pm 17$ \\
2026-04-28 & 11.5 & 0.41 & 0.15 & $-$16 & 21 & $80 \pm 21$ \\
\end{longtable}
\endgroup

\bibliography{main}{}
\bibliographystyle{aasjournalv7}

\end{document}